\documentstyle[epsfig]{aa}
\begin{document}
   \title{The emissivity of dust grains in spiral galaxies}

   \subtitle{}

\author{P.B. Alton \inst{1} 
\and E.M. Xilouris\inst{2} 
\and A. Misiriotis\inst{3} 
\and K.M. Dasyra\inst{4}
\and M. Dumke\inst{5}}
	

   \offprints{E.M. Xilouris: xilouris@astro.noa.gr}

\institute{Le Mas Fleuri, 44, Chemin des Guiols, 06650 Le Rouret, France.
         \and
             National Observatory of Athens, I. Metaxa \& Vas. Pavlou str., Palaia Penteli, GR-15236, Athens, Greece.
	\and
           University of Crete, Physics Department, PO Box 2208, 71003 Heraklion, Crete, Greece.
	\and
	   Max-Planck-Institut f\"ur Extraterrestrische Physik, Postfach 1312, D-85741 Garching, Germany. 
	\and
	   Max-Planck-Institut f\"ur Radioastronomie, Auf dem H\"ugel 69, 53121 Bonn, Germany.
             }

   \date{}

   \abstract{
We use the radiation transfer simulation of Xilouris et al. (\cite{xilouris1999}) to constrain the quantity 
of dust in three nearby spiral galaxies (NGC 4013, NGC 5907 and NGC 4565). The predicted visual optical depth 
from the model is compared with the thermal continuum radiation detected from NGC 4013 and NGC 5907 at 
$850~{\mu}m$ and from NGC 4565 at 1.2 mm. The former is based on SCUBA images of NGC 4013 and NGC 5907, 
reduced and presented for the first time in this work. The comparison of visual optical depth and $850~\mu$m 
(1.2 mm) emission yields the emissivity of dust grains in the submillimeter (millimeter) waveband. We infer a 
value of $1.2 \times 10^{4}$ for the emissivity at $850~\mu$m which is a factor 4 higher than the benchmark, 
semi-empirical model of Draine \& Lee (\cite{draine1984}). At 1.2 mm our values are a factor 1.5 higher than 
this model. Our estimates are more closely aligned with recent measurements carried out in the laboratory on 
{\em amorphous} carbon and silicate particulates. A comparison between the distribution of $850~\mu$m (1.2 mm) 
surface brightness and the intensity levels in the  $^{12}$CO(1-0) and 21 cm lines underlines the spatial 
association between dust detected in the submillimeter/millimeter waveband and molecular gas clouds. We 
suggest that the relatively high emissivity values that we derive may be attributable to amorphous, fluffy 
grains situated in denser gas environments.

              
   }

   \maketitle
%

\section{Introduction}
\label{intro}

The continuing improvement in Far-Infrared (FIR) instrumentation has permitted a more complete spectral 
sampling of the thermal continuum radiation emitted by dust grains situated in external spiral disks. The 
result has been that the FIR optical depth of some nearby galaxies is now constrained to within a factor 2-3 
(Dupac et al. \cite{dupac2003}). Ironically, the grain parameter allowing the FIR optical depth to be converted 
into dust column density and ultimately dust mass -- the emissivity $Q$ -- still remains uncertain by nearly an
 order of magnitude (Hughes et al. \cite{hughes1993}). $Q$ indicates the efficiency with which dust grains of 
 a particular temperature emit FIR thermal radiation (sometimes this property is expressed as the mass absorption
  coefficient $\kappa$; both $Q$ and $\kappa$ will be defined formally in Sect.~\ref{tech}). Due to a dearth of
   direct measurements, a large proportion of submillimeter (submm) and millimeter (mm) astronomers rely on a 
   single estimate of emissivity at $125~{\mu}m$ obtained 20 years ago for a single Galactic reflection nebula 
   (Hildebrand \cite{hildebrand1983}). Given, as we shall see, that emissivity varies both strongly with 
   wavelength ($\lambda$) and environment, $Q$ can be predicted to within little better than an order of 
   magnitude at wavelengths close to 1 mm. Another work frequently cited is that of Draine \& Lee 
   (\cite{draine1984}). These authors propose emissivities on the basis of laboratory experiments for 
   $\lambda \leq 60~{\mu}m$ and use primarily a model based on solid state theory to extrapolate to longer 
   wavelengths. The $100~{\mu}m$-brightness per H-atom, predicted by Draine \& Lee (\cite{draine1984}) for 
   high-lattitude dust clouds, agrees within 30\% of the measurements carried out by the 
   {\it Infrared Astronomical Satellite} (IRAS). However, as the authors themselves acknowledge 
   (p.107 of Draine \& Lee \cite{draine1984}), the emissivity proposed for $\lambda > 300~{\mu}m$ may be too 
   low by a factor of 3-4 to be consistent with astronomical observations.

In this paper, we carry out a direct measurement of the submm/mm emissivity of dust grains situated in three 
nearby, quiescent spiral disks -- NGC 4013, NGC 5907 and NGC 4565. The technique, explained in detail in
 Sect.~\ref{tech}, compares two tracers of what is believed to be the same dust grain population in order to 
 infer the emissivity. The tracers in question are submm/mm surface brightness (Sect.~\ref{submm}) and visual 
 optical depth (Sect.~\ref{tau_v}). We have already undertaken a similar analysis for the nearby spiral NGC 891 
 and, in this case, our results imply an $850~{\mu}m$ emissivity which is 3 times higher than the Draine \& Lee 
 (\cite{draine1984}) value (Alton et al. \cite{alton2000}). In Sect.~\ref{discussion}, we compare our results 
 with estimates of emissivity in the literature for grains situated in various astrophysical environments. Apart  
from estimating the submm/mm emissivity we explore how the dust traced in thermal continuum emission relates to 
the various gas phases (H$_{2}$, HI) within the disk (Sect.~\ref{gas}). 


\section{Technique for determining the emissivity}
\label{tech}

Alton et al. (\cite{alton2000}) have already discussed the method in some detail but we summarise here the basic
 precepts. Classical dust grains of radius $0.1~{\mu}m$ will tend to reach an equilibrium temperature $T$ when 
 immersed in a stellar radiation field. Under such circumstances, heating due to the absorption of optical, 
 ultraviolet and near-infrared photons is exactly balanced by cooling due to emission of mid and far-infrared 
 radiation. The efficiency with which grains emit submm/mm radiation depends on their composition and structure 
 and can be expressed by the emissivity $Q$, a dimensionless quantity which indicates how the flux density ($F$) 
 recorded at wavelength $\lambda$ compares with that emitted by a blackbody:

\begin{equation}
\label{eq_fd}
F(\lambda) = \frac{n\sigma}{D^{2}} Q(\lambda)B(\lambda,T)
\end{equation}

\noindent Here $D$ is the distance to a dust cloud containing $n$ dust grains of geometrical cross-section $\sigma$. $B$ is the planck function for a blackbody of temperature $T$.\footnote{A related quantity is the dust absorption coefficient $\kappa$ which, for grains of radius $a$ and material density $\rho$, is related to $Q$ as follows: $\kappa = \frac{3Q}{4a\rho}$.}

In the past, a major obstacle has been constraining $B(\lambda,T)$ which, near the peak of the FIR spectrum ($\lambda \sim 200~{\mu}m$ for spiral galaxies), is highly sensitive to the adopted grain temperature $T$. The temperatures we shall use, however, are derived from fitting the Rayleigh-Jeans tail of spectrum and in this regime $B \propto T$ approximately. We shall argue that moderate errors in $T$ due to the fitting process ($\simeq$30\%) are {\em not} the major source of uncertainty in deriving $Q$. Inspecting Eq.~(\ref{eq_fd}), we recognise that, even if $F$ and $B$ are known with relative certainty we are required to disentangle $n\sigma$ from $Q$. We achieve this by substituting the visual optical depth defined as follows:

\begin{equation}
\label{eq_tau}
\tau_{V} = N \sigma Q(V)
\end{equation}

\noindent where $Q(V)$ is the extinction efficiency in the V-band ($\lambda = 0.55~{\mu}$m) and $N$ is the number of grains per unit area on the sky. At the same time we substitute the surface brightness $f(\lambda)$ for the flux density $F(\lambda)$ so that the distance to the dust cloud $D$ is eliminated. This re-formulation of Eq.~(\ref{eq_fd}) yields:

\begin{equation}
\label{eq_sb}
\frac{\tau_{V}}{f(\lambda)} = \frac{Q(V)}{Q(\lambda)} \frac{2.2 \times 10^{-18}}{B(\lambda,T)}
\end{equation}

\noindent where $f(\lambda)$, for reasons apparent later, is expressed as Jy/$16''$~beam.

In general, it is the ratio $\frac{Q(V)}{Q(\lambda)}$ which is of immediate interest in comparing our results with other studies. However, if we are to quantify the number of FIR-emitting grains in Eq.~(\ref{eq_fd}), and thereby ultimately determine the dust mass of the system, we should derive the absolute emissivity $Q(\lambda)$. To do this we must specify $Q(V)$, $B(\lambda,T)$, $f(\lambda)$ and $\tau_{V}$ in Eq.~(\ref{eq_sb}). The first of these quantities, $Q(V)$, is believed to lie, with some certainty between 1 and 2 (Whittet \cite{whittet1992}, (p. 60); Alton \cite{alton1996}) and we assume a value of 1.5 hereafter. The temperature associated with the blackbody $B(\lambda,T)$ follows from the shape of the FIR spectrum. The submm/mm surface brightness, $f(\lambda)$, and the optical depth $\tau_{V}$ are the two quantities which we shall bring together in this paper in order to determine $Q(\lambda)$. We start by discussing the submm/mm surface brightness and in the subsequent section we show how $\tau_{V}$ is derived.

\begin{figure}
\centering
\epsfig{figure=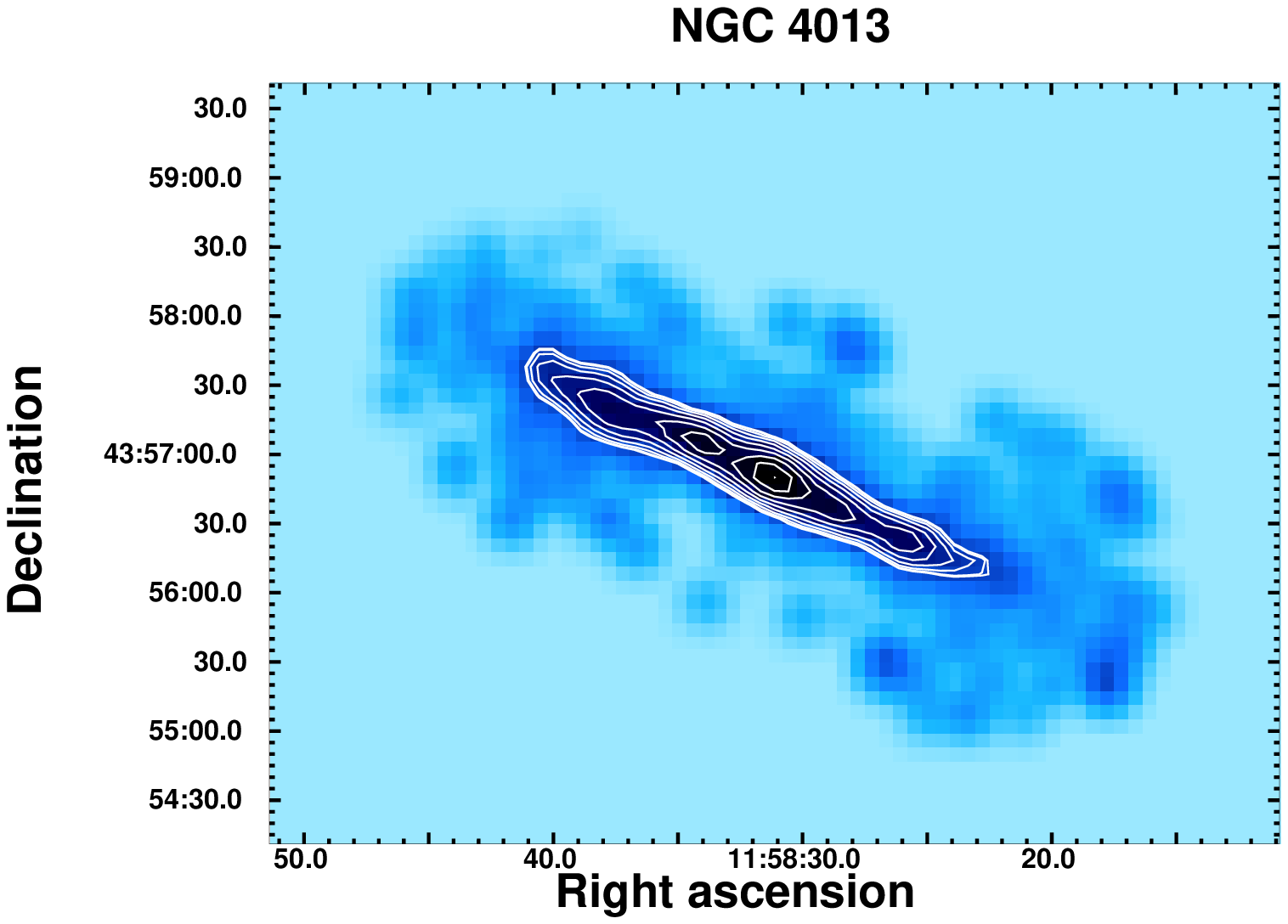,width=10.5cm}
\epsfig{figure=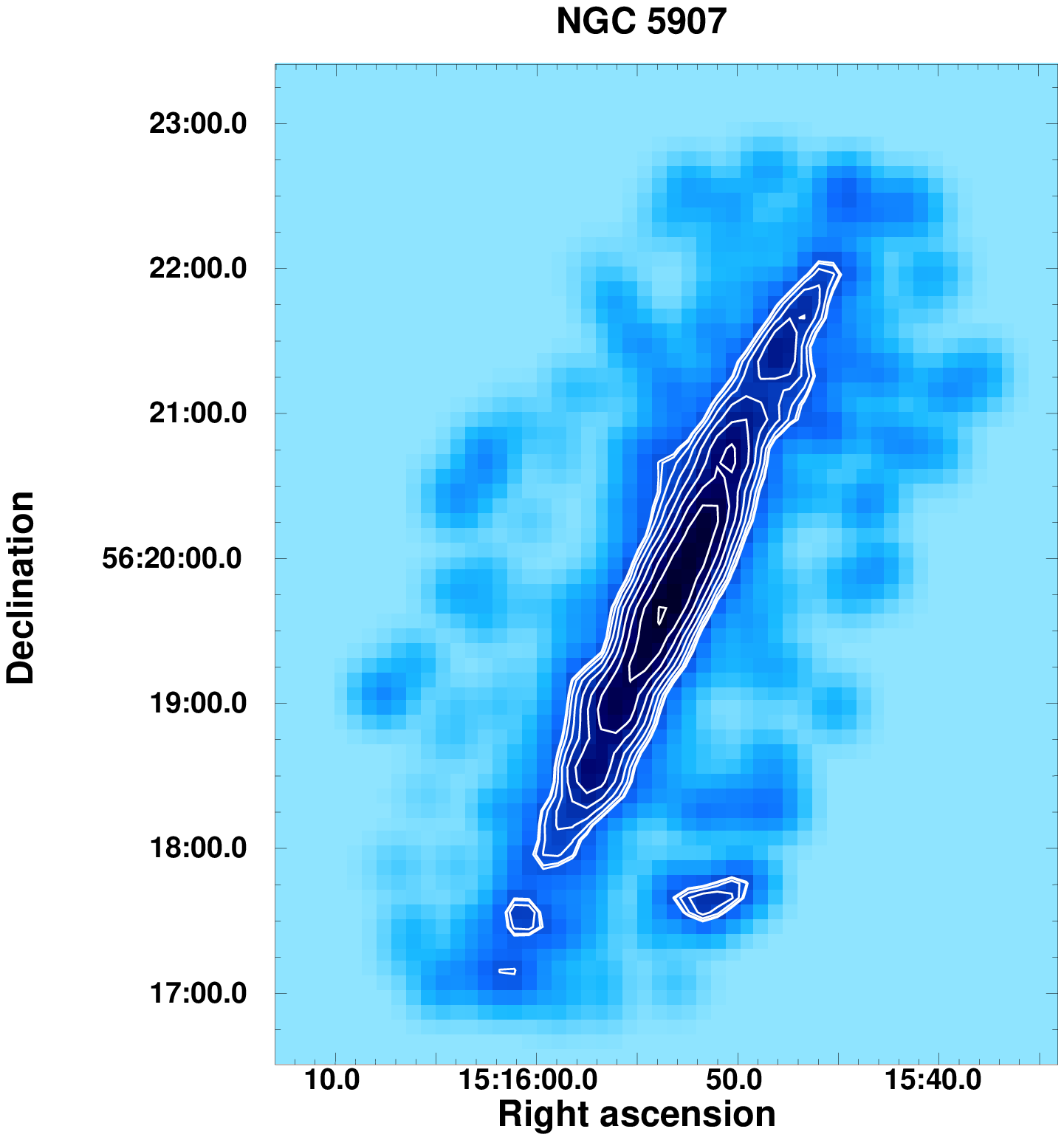,width=10.5cm}
\caption{NGC 4013 (top) and NGC 5907 (bottom) observed in the $850~\mu$m continuum with SCUBA.
Both the background image and the contours refer to emission at $850~\mu$m.
For NGC 4013 the contour levels are at 10.0, 10.4, 11.4, 13.3 15.9, 19.2, 23.3,
28.1, 33.7 and 40.0
mJy/$16''$~beam  while for
NGC 5907 the contour levels are at 20.0, 20.8, 23.4, 27.8, 33.8, 41.6, 51.1, 62.3, 75.3 and
90.0 mJy/$16''$~beam. In both cases the first contour level is at the 3 $\sigma$
level and the beam size is $16''$(FWHM).}
\label{fig_scuba}
\end{figure}

\begin{figure}
\epsfig{figure=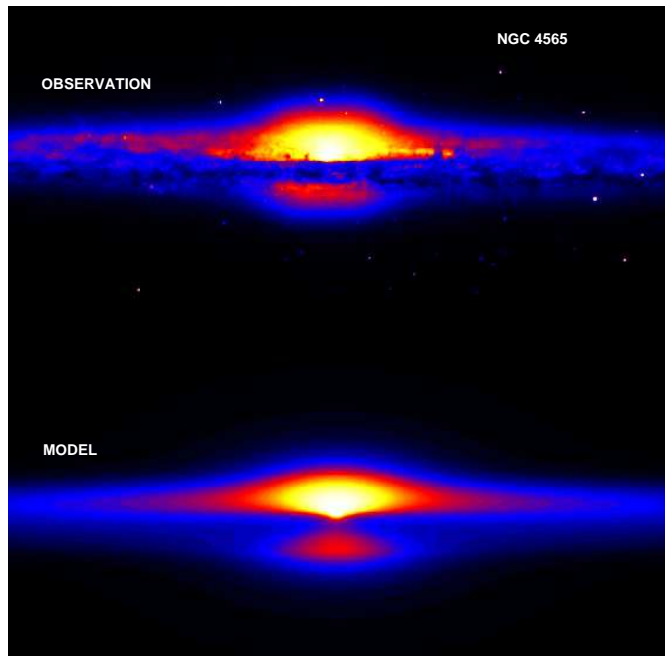,width=8.8cm,angle=0}
\caption{Radiation transfer modelling of NGC 4565 in the V-band. The real galaxy,
as observed by Howk \& Savage (\cite{howk1999}), is shown at the top whilst the corresponding image generated
by our radiation transfer simulation is situated at the bottom.}
\label{fig_composite}
\end{figure}
                                                                                                            
\section{Submillimeter/Millimeter Data}
\label{submm}

Our submm/mm data originate from the following sources. The archive of the James Clerk Maxwell Telescope (JCMT) 
yielded 450 and $850~\mu$m exposures of both NGC 4013 and NGC 5907. These raw images had been taken with the 
Submillimeter Common User Bolometer Array (SCUBA) and have never been published previously. A reduced and 
fully-cleaned image of NGC 5907 at 1.2 mm, taken with the bolometer array at IRAM (Institut de Radioastronomie 
millimetrique), was provided by Dumke et al.
(1997). For NGC 4565, we extracted the IRAM major axis profile at 1.2 mm which appears in Neininger et al. 
(\cite{neininger1996}).

\begin{table}
\caption[]{Best-fit parameters derived from RT-modelling of NGC 4013, NGC 5907 and NGC 4565 in the V-band.
$\tau_{V}^{f}$ is the visual optical depth when the galaxy is viewed face-on and $\theta$ is the inferred
inclination of the object. $h_{d}$ and $z_{d}$ are, respectively, the exponential scalelength and
exponential scaleheight of the dust assuming the distance listed in the second column. Distances to NGC 4013, NGC 4565
and NGC 5907 are taken, respectively, from Bottema (1995), Rupen (1991) and van der Kruit \& Searle (1982).  }
\begin{tabular}{cccccc}
\hline
\hline
Galaxy    & Distance & $\tau_{V}^{f}$ & $h_{d}$ & $z_{d}$ & $\theta$  \\
          &   (Mpc)  &                &  (kpc)  &  (kpc)  &   (deg.)  \\ \hline
NGC 4013  &    12    &   0.67         &   2.45  &   0.13  &   89.7    \\
NGC 4565  &    10    &   0.63         &   9.7   &   0.21  &   88.0    \\
NGC 5907  &    11    &   0.49         &   5.29  &   0.11  &   87.2    \\ \hline
\end{tabular}
\label{tab_model}
\end{table}

SCUBA is mounted at the Nasmyth focus of JCMT and provides simultaneous imaging at 450 and $850~{\mu}m$ for a 
region of sky $2.3'$ in diameter (Holland et al. \cite{holland1999}). The shortwave array ($450~{\mu}m$) consists
 of 91 bolometers (HPBW = $7.5''$) whilst the longwave array ($850~{\mu}m$) is composed of 37 elements 
 (HPBW = $14.7''$). In order to provide fully sampled images the secondary mirror moves in a 64-step jiggle 
 pattern with the integration time lasting 1 sec at each position. At the same time the secondary mirror chops 
 at 7 Hz so as to remove the reference (sky) emission. After the 16 steps of jiggle pattern, the telescope nods 
 in order to allow for slowly varying sky gradients. For both NGC 4013 and NGC 5907 the chop throw was $120''$ 
 perpendicular to the major axis. Hourly `sky-dip' measurements allow the sky opacity to be monitored during the
  course of the eight-hour observing shift.

The dedicated SCUBA software package, SURF (Jenness et al. \cite{jenness1997}), was used to clean, flat-field 
and calibrate the raw images of NGC 4013 and NGC 5907 according to atmospheric attenuation. SURF also 
facilitates a reduction in the final image noise by compensating for spatially-correlated sky emission across 
the field of view. The pointing stability of the observations was better than $5''$ and the typical $850~\mu$m 
zenith opacity was 0.16 for NGC 4013 and 0.34 for NGC 5907.

To complement the SCUBA and IRAM images we obtained resolution-enhanced IRAS images of NGC 4013, NGC 5907 and 
NGC 4565 at 60 and $100~{\mu}m$ (HiRes images made available by IPAC). The mean spatial resolution of these data 
was $\simeq 1'$ at $60~{\mu}m$ and $\simeq 1.5'$ at $100~{\mu}m$. The SCUBA $850~\mu$m images for NGC 4013 and
 NGC 5907 are shown in Fig.~\ref{fig_scuba}. The corresponding instrumental beam  is $\simeq 16''$ (FWHM). The 
 SCUBA $450~\mu$m data are much noisier than the longwave measurements and are not presented here. For NGC 4013,
  we were still in a position to obtain a global flux density at $450~\mu$m using the precepts outlined in 
  Sect.~\ref{sed}.

For both NGC 4013 and NGC 5907 the $850~\mu$m morphology is fairly clumped. NGC 4013 possesses an asymmetric 
distribution with respect to the nucleus with the north-east side of the disk distinctly brighter. An 
enhancement in $850~\mu$m emission from one side of the disk was a feature noted in the morphology of NGC 891 
(Alton et al. \cite{alton1998}). Such `hotspots' cannot be attibuted to viewing effects (e.g. a spiral arm 
located on the near-side) since the emission in the FIR is expected to be optically-thin.

\begin{table}
\caption[]{Global flux densities, as defined in Sect.~\ref{sed} for NGC 4013, NGC 5907 and NGC 4565. 
For each galaxy, the flux density $F_{\lambda}$ is given in Jy at wavelength ${\lambda}$ (in microns).
The uncertainty in the flux density is $\pm$ 15\%.}
\begin{tabular}{lccccc}
\hline
\hline
Galaxy & F$_{60}$ & F$_{100}$ & F$_{450}$ & F$_{850}$ & F$_{1200}$ \\
       &  (Jy)         & (Jy)           & (Jy)           & (Jy)        & (Jy)   \\
\hline
NGC 4013 & 7.0       & 23           & 4.1          & 0.61      &  --  \\
NGC 4565 & 7.1       & 29           & --           &  --       & 0.55  \\
NGC 5907 & 16        & 56           & --           & 1.6       & 0.54 \\
\hline
\end{tabular}
\label{tab_fluxes}
\end{table}
\begin{table}
\caption[]{Dust components derived from SED fitting of NGC 4013, NGC 5907 and NGC 4565 in the FIR/submm/mm
regime. We decompose the spectrum into 2 components; warm dust of temperature $T^{W}$ and cold dust of 
temperature $T^{C}$. The ratio of cold dust mass to warm dust mass is given as $\frac{M^{C}}{M^{W}}$. An index of 1.5 has 
been assumed for the wavelength dependency of the emissivity ($Q\propto \lambda^{-1.5}$).}
\begin{tabular}{cccc}
\hline
\hline
Galaxy    &    $T^{W}$ & $T^{C}$ &  $\frac{M^{C}}{M^{W}}$  \\
          &     (K)    &  (K)    &                         \\ \hline
NGC 4013  &      39    &  23     &         49              \\
NGC 4565  &      43    &  20     &         99              \\
NGC 5907  &      28    &  13     &         5.7             \\ \hline
\end{tabular}
\label{tab_sed}
\end{table}

\section{Derivation of visual optical depth}
\label{tau_v}

Maps of visual optical depth for NGC 4013, NGC 5907 and NGC 4565 were derived using the Radiation Transfer (RT)
 model of Xilouris et al. (\cite{xilouris1999}). This numerical model simulates the absorption and scattering of
  optical and near-infrared photons by dust grains situated in a stellar disk using the Henyey-Greenstein phase 
  function and the Galactic albedo (Lillie \& Witt \cite{lillie1976}). Dust and stars are assumed to be mixed 
  and smoothly distributed. Both components behave independently, falling off as exponential distributions in 
  both z-height and radius. The bulge is treated as a dust-free component containing stars decreasing as 
  $R^{\frac{1}{4}}$. The scalelength, scaleheight and optical depth of the exponential dust disk are adjusted 
  until there is a match between {\em predicted} surface brightness and {\em observed} surface brightness over
   the entire edge-on disk.

\begin{figure}
\centering
\epsfig{figure=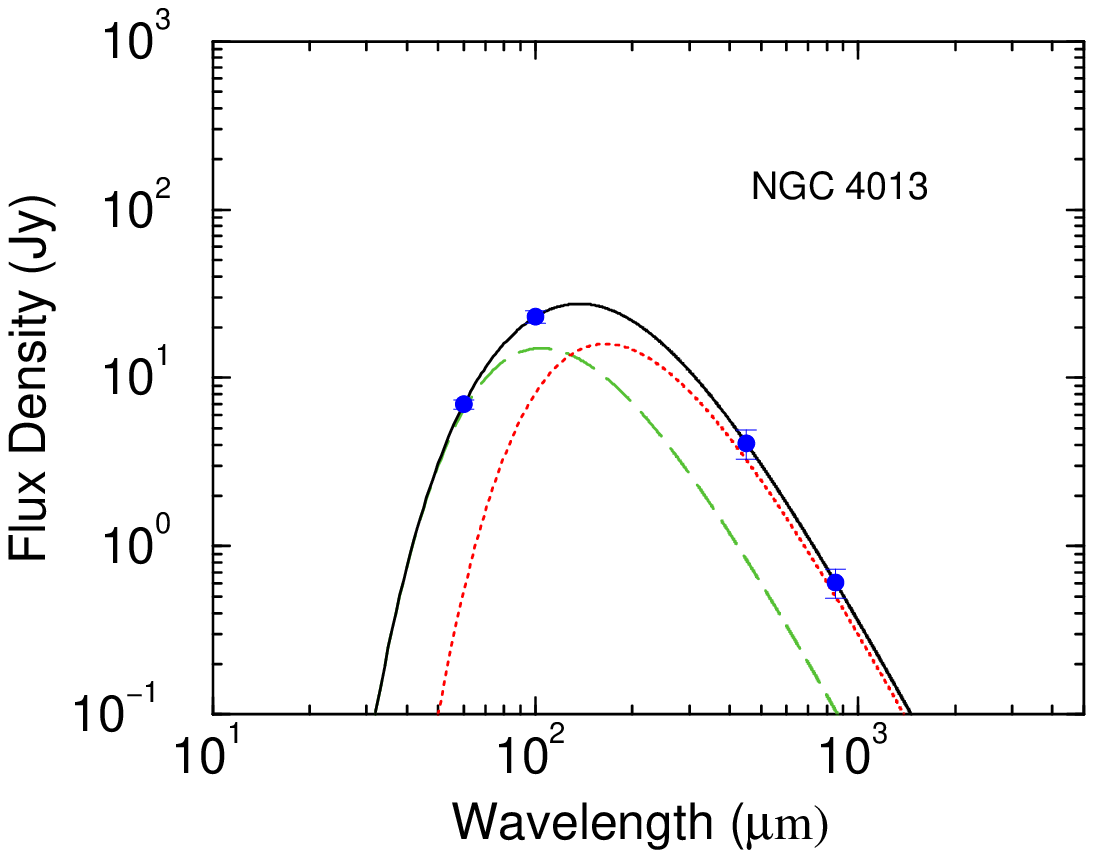,width=8.5cm}
\epsfig{figure=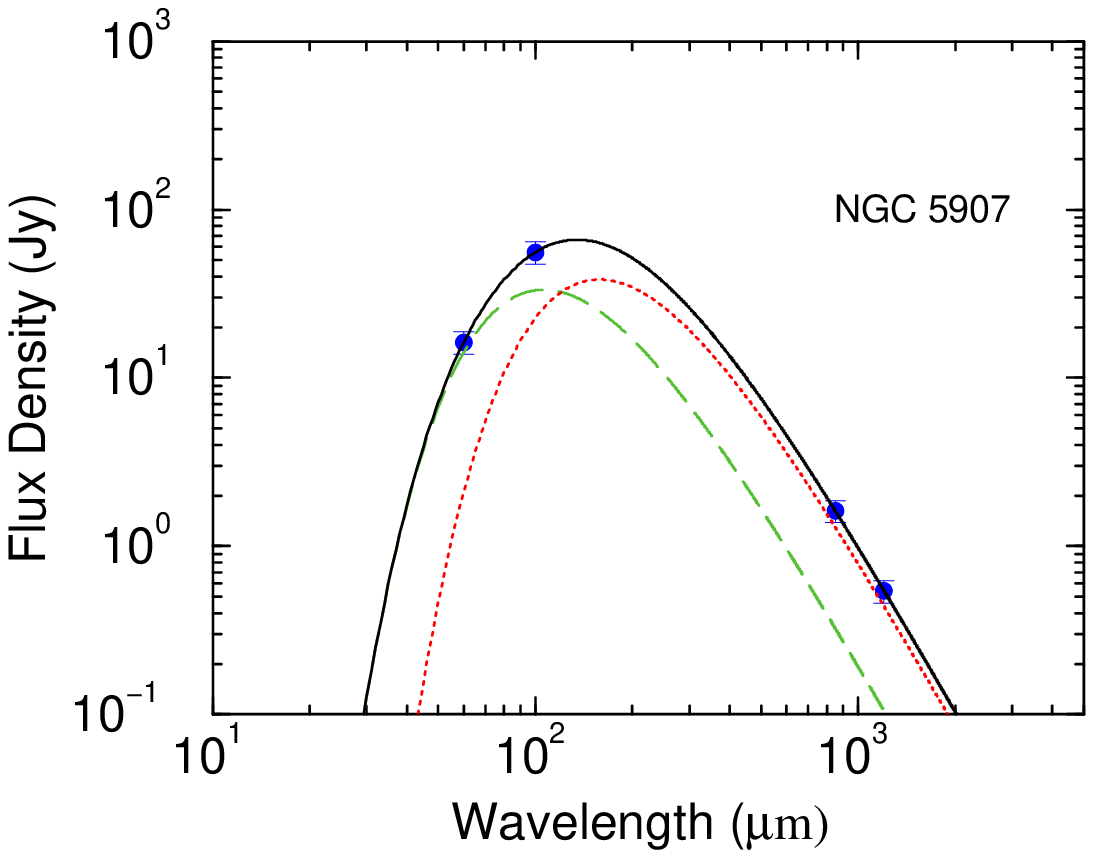,width=8.5cm}
\epsfig{figure=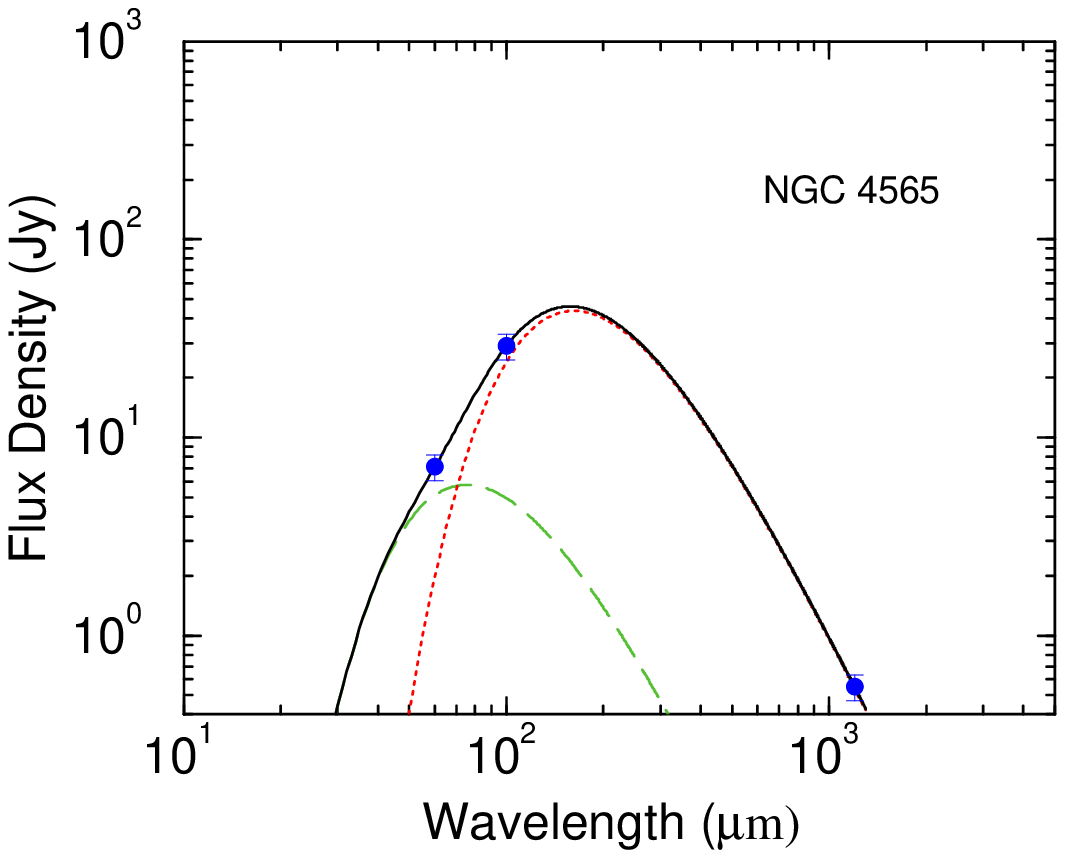,width=8.5cm}
\caption{Flux densities for NGC 4013, NGC 5907 and NGC 4565 as observed by IRAS
(60 and $100~\mu$m), SCUBA (450 and $850~\mu$m) and IRAM (1.2 mm). In each case, the thermal
spectrum has been fitted by the sum of a warm and cold dust component (dashed and dotted
lines respectively). The total SED is given by the solid curve. The proportion of dust in
each of the warm and cold components, as well as their associated temperatures,
are given in Table \ref{tab_sed}.}
\label{fig_sed}
\end{figure}

Our method is best suited to galaxies lying within $5^{\circ}$ of edge-on (hence the current sample of 
highly-inclined galaxies). This flat orientation allows us to: (1) approximate, within a limited extent, a 
clumpy dust disk by a smooth exponential distribution and (2) use the unattenuated surface brightness at large
 z-heights as a boundary condition for fitting the dust distribution (see Fig.~1 in Xilouris et al. 
 \cite{xilouris1997}). Although the model possesses certain limitations (to be discussed in 
 Sect.~\ref{underestimate}), the opacities inferred independently in several wavebands (B, V, R, I, J, K)
  are consistent with a Milky Way extinction law within a few percent giving us a fairly high degree of 
  confidence in the simulation (see Fig.~12 of Xilouris et al. \cite{xilouris1999}).

The RT simulations for NGC 4013 and NGC 5907 have already been presented in Xilouris et al. 
(\cite{xilouris1999}) and we adopted the corresponding dust parameters without further treatment. 
The simulation for NGC 4565 was new and required at least one optical/NIR image for the fitting procedure.
 To this end, a reduced and cleaned V-band image of NGC 4565 was retrieved from the NED server (originally 
 taken by Howk \& Savage (\cite{howk1999}) using the WIYN telescope).  Fig.~\ref{fig_composite} illustrates
  both the observed and simulated images for this object. Table \ref{tab_model} lists the model parameters, 
  including the visual optical depth, for all three galaxies in the sample. As indicated in this table, we 
  adopt distances of 12, 10 and 11 Mpc to NGC 4013, NGC 4565 and NGC 5907, respectively.

\section{Grain temperature from SED fitting}
\label{sed}
 
The spectral energy distribution (SED) for NGC 4013 and NGC 5907 was derived in the following manner. The SCUBA, IRAM
and HiRes $60~\mu$m images were smoothed to a common `worst' spatial resolution namely that of the $100~\mu$m 
IRAS image ($\simeq 1.5'$ FWHM). To avoid aperture corrections (Chini et al 1995; Hughes et al 1997)
 an image area was defined over which the object had been detected in all 
FIR/submm/mm filters. The emission within this area was used as the global flux density. The arcminute 
resolution of the IRAS data is too coarse to monitor the change in grain temperature with galactocentric radius
 and, therefore, we sought only to fit global mean dust temperatures to the thermal spectrum of each object. For
  NGC 4565 we integrated the major axis profile at 1.2 mm appearing in Neininger et al. (1996) and measured the 
  60 and $100~\mu$m flux densities over the same area of our HiRes images. The global flux densities we derive 
  for all three objects are shown in Table~\ref{tab_fluxes} and the corresponding SEDs are depicted in 
  Fig.~\ref{fig_sed}.

\begin{figure}
\centering
\epsfig{figure=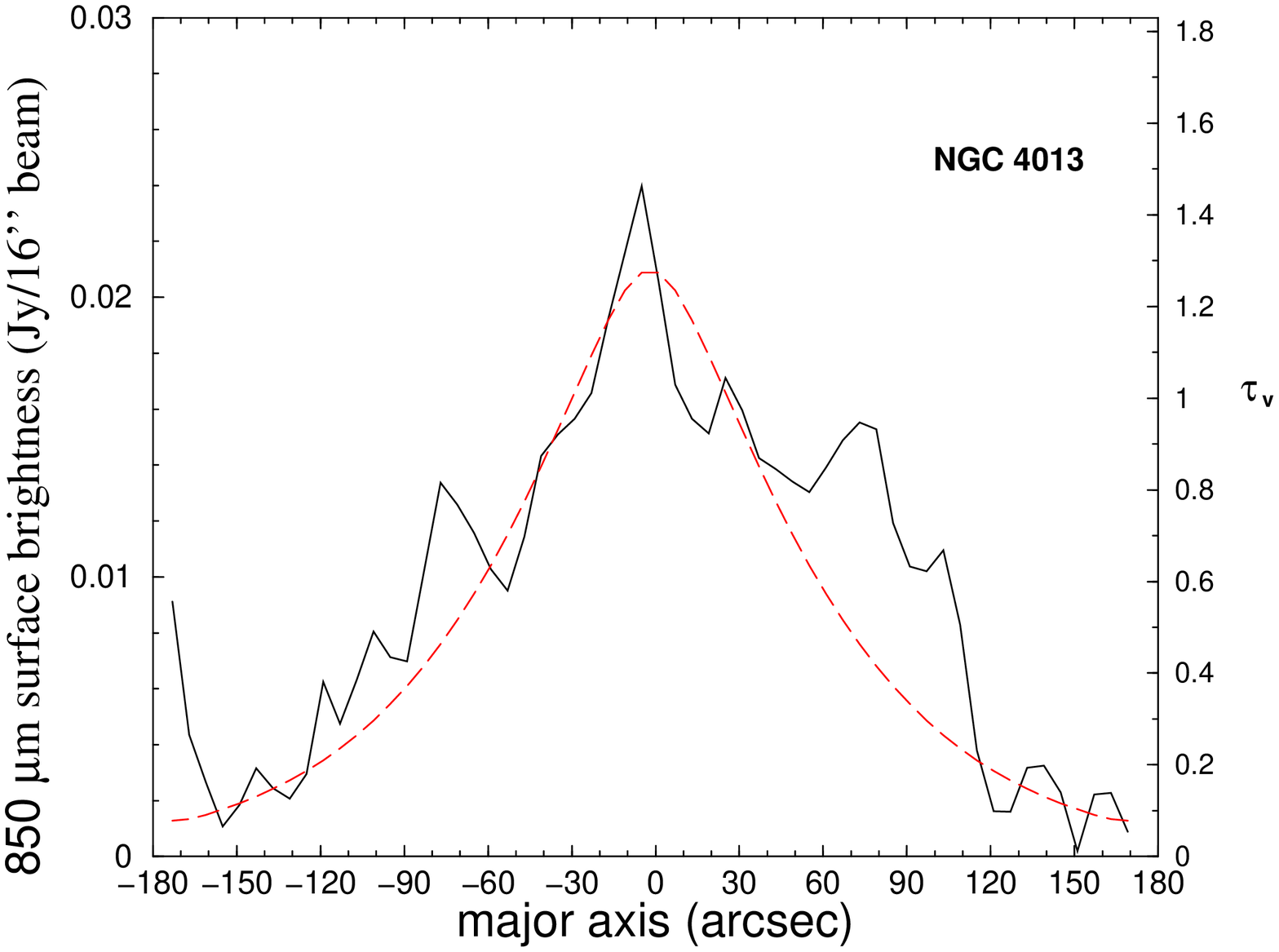,width=8.5cm}
\epsfig{figure=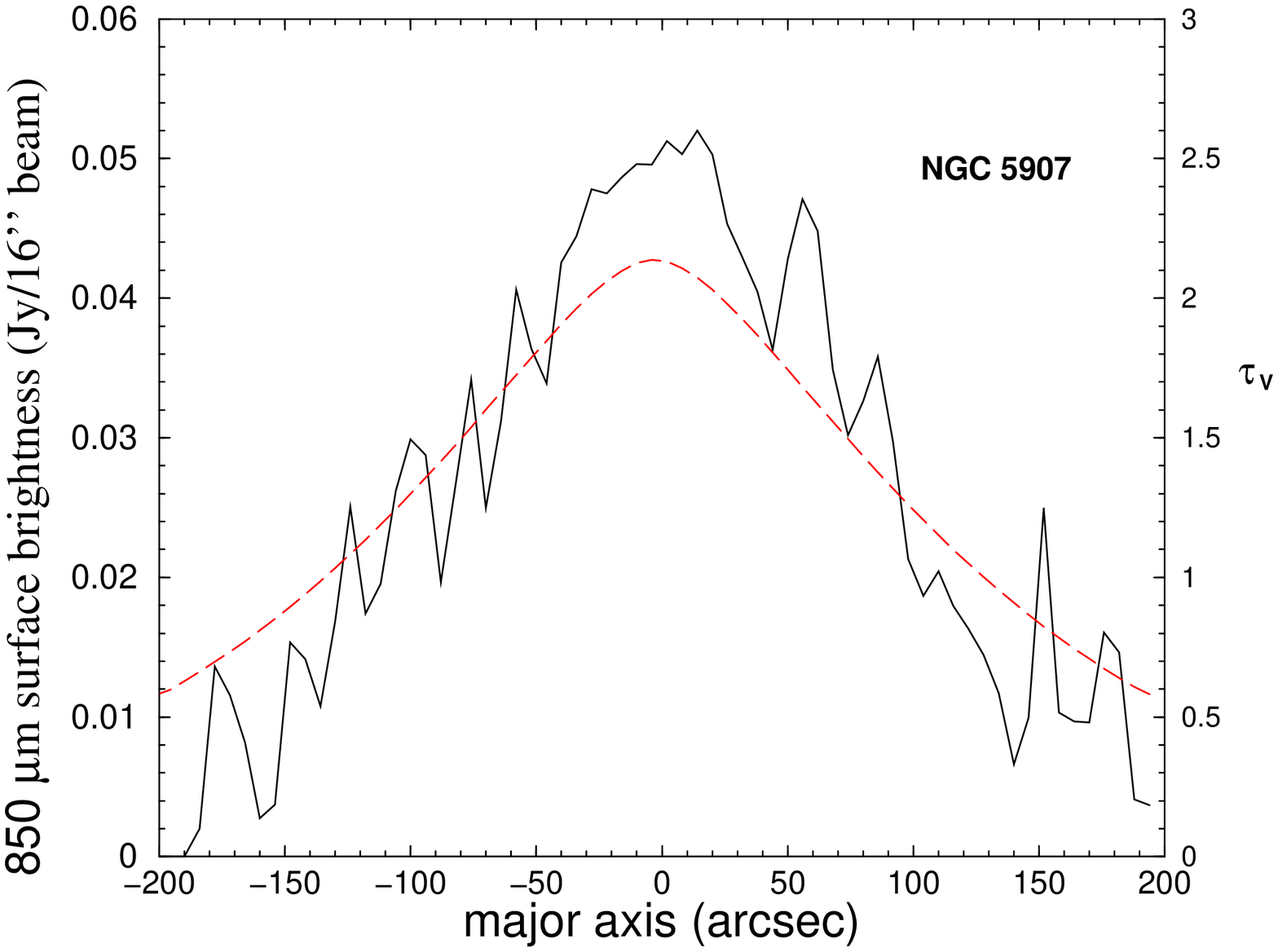,width=8.5cm}
\epsfig{figure=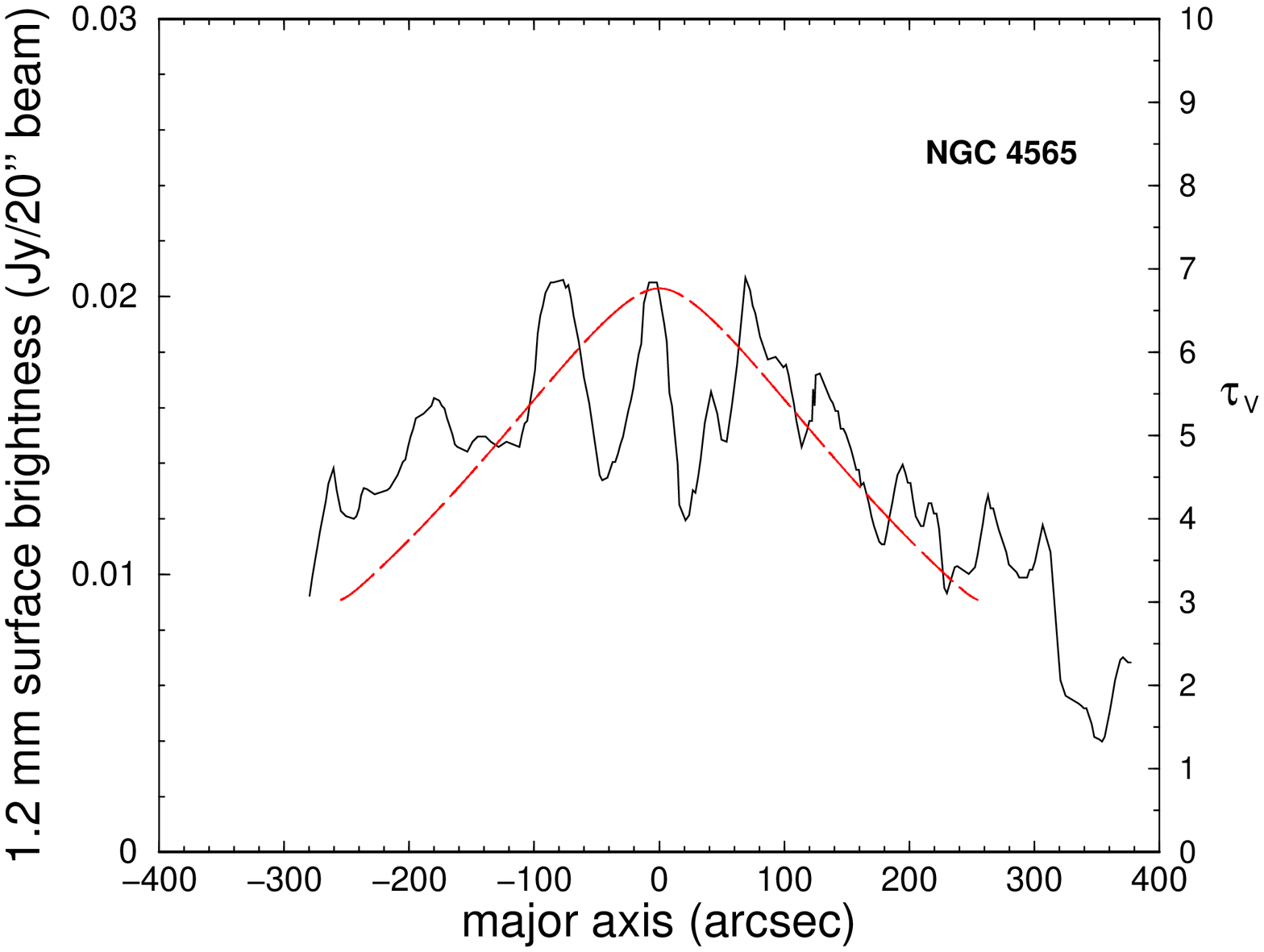,width=8.5cm}
\caption{Major axis profiles of NGC 4013, NGC 5907 and NGC 4565 in $850~\mu$m and 1.2 mm surface
brightness (solid line, left axis) and visual optical depth (dashed line, right
axis).
The latter has been inferred from radiation transfer modelling and the map in visual
optical depth has been smoothed to the corresponding spatial resolution of the submm/mm
data before profiling ($16''$ and $20''$ at $850~\mu$m and 1.2 mm respectively).
The major axis is defined such that positive values correspond to north-east, south-east and north-west
for NGC 4013, NGC 5907 and NGC 4565 respectively.}
\label{fig_prof}
\end{figure}

To determine the grain temperature we assume that the flux densities recorded in Fig.~\ref{fig_sed} arise purely 
from thermal continuum emission. Contamination of the 450 and $850~\mu$m filters by the respective $^{12}$CO(6-5)
 and $^{12}$CO(3-2) line transitions was neglected in the first instance. Similarly, no account was taken at 
 this stage of non-equilibrium emission from small grains. Thus, we decomposed each SED into two dust components 
 (warm and cold) assuming that each component emits as a modified blackbody with 
 $\beta = 1.5$ [$F(\lambda,T)\propto B(\lambda,T) {\lambda}^{-{\beta}}$]. Our reasons for chosing this 
 relatively low value of $\beta$ will become evident at a later stage in the paper 
 (Sect.~\ref{sed_uncertainties}). The warm and cold grain temperatures derived from the fitting procedure are 
 summarized in Table~\ref{tab_sed}. We emphasize that our aim here is to fit the {\em shape} of the SED. 
 We are not seeking to derive absolute dust masses which would require adopting {\em a priori} a FIR emissivity. 
 We derive 3 parameters (cold dust temperature, warm dust temperature and ratio of cold to warm dust mass) 
 from either 3 (NGC 4565) or 4 data points (NGC 4013, NGC 5907). 

From Table~\ref{tab_sed}, we surmise that 85-99\% of the dust detected is cold possessing a temperature of 
13-23 K. Similar values are recorded for diffuse Galactic dust mapped with COBE (Reach et al. \cite{reach1995}) 
and from authors carrying out more complete spectral mapping of external galaxies (e.g.~Dupac et al.
 \cite{dupac2003}). A discussion of the uncertainties in the grain temperature and its implications for the inferred emissivity is reserved to Sect. 7.2.

\begin{table*}
\caption[]{Values for submillimeter and millimeter emissivity, $Q(\lambda)$, collated from the literature.
Estimates are given as $\frac{Q(V)}{Q(\lambda)}$ where the extinction efficiency in the V-band [$Q(V)$]
 has been taken, where necessary, to be 1.5 (Whittet 1992). The absorption mass coefficient $\kappa(\lambda)$ is
calculated from $Q(\lambda)$ assuming a classical grain radius of $0.1~\mu$m and a material density of
$\rho =3 \times 10^{3}~kgm^{-3}$. For each measurement the column `Medium' denotes the type of astrophysical
environment where the corresponding grains are believed to be situated.}
\begin{tabular}{llllll}
\hline
\hline
$\lambda$ & $\frac{Q(V)}{Q(\lambda)}$ & $\kappa(\lambda)$ & Medium &
 Reference & Comment \\
($\mu$m)  &  & ($cm^{2} g^{-1}$) & & & \\ \hline
100 & 760 & 49 & diffuse HI (Milky Way) & Bianchi et al. (1999)       & \\
125 & 2000 & 19 & reflection nebula & Hildebrand (1983)    & $\frac{Q(V)}{Q(\lambda)}$ up to 5000 if $\beta=1$ \\
240 & 3900 & 9.7 & diffuse HI (Milky Way) & Boulanger et al. (1996) & DIRBE data; values also given \\
    &      &     &                  &                                      & at 100 and 140$~\mu$m \\
250 & 2500 & 15  & reflection nebula & Casey (1991)                 & 5 objects; $\frac{Q(V)}{Q(\lambda)}$ up to 30000  \\
    &      &     &                   &                              &   depending on the method   \\
250 & 4400 & 8.5 & diffuse HI (model) & Draine \& Lee (1984)   & Uses laboratory data; oscillator  \\
    &      &     &                    &                                    & model for $\lambda>60~\mu$m \\
400 & 1900 & 20 & late-type star & Sopka (1985)                 &    \\
736 & 34000 & 1.1 & diffuse HI (Milky Way) & Boulanger et al. (1996) & FIRAS data; values also given \\
    &       &     &                 &                                      & at 346, 490, 535 and 1100$~\mu$m \\
850 & 9100 & 4.1 & cold core        & Rengarajan (1984)        &  \\
850 & 30000 & 1.3 & cold core       & Kramer et al. (2003)          &  \\
850 & 54000 & 0.69 & diffuse HI ?   & James et al. (2002)        & Using $N_{H}$, metallicity \& SCUBA data \\
850 & 16000 & 2.4 & H$_{2}$ (NGC 891) & Alton et al. (2000)        & RT model of this work \& SCUBA data \\
850 & 14000 & 3.2 & H$_{2}$ (NGC 4013) & this work                           &  \\
850 & 12000 & 2.7 & H$_{2}$ (NGC 5907) & this work                           &  \\
850 & 25000 & 1.5 & cold core       & Bianchi et al. (2003)    & SCUBA data\ \\
1000 & 18000 & 2.1 & diffuse HI (model) & Mathis \& Whiffen (1989)  & \\
1000 & 180000 & 0.22 & laboratory (Si crystalline) & Agladze et al. (1996);  & Mean value of both works \\
     &       &       &                             & Mennella et al. (1996)  & (150000 and 197000) \\
1000 & 40000 & 0.95 & laboratory (Si amorphous) &  Agladze et al. (1996);    & Mean value of both works \\
     &       &       &                          &  Mennella et al. (1996)    &  (36100 and 43600) \\
1000 & 1600   & 24    & laboratory (C amorphous)   & Mennella et al. (1996)  &                  \\
1200 & 120000 & 0.315 & diffuse HI (NGC 4565)      & Neininger et al. (1996) & Authors claim emissivity 2-4 \\
     &        &       &                            &           &   higher for molecular regions \\
1200 & 110000 & 0.35 & diffuse HI (NGC 5907) & Dumke et al. (1997) & Outer regions of the galactic disk \\
1200 & 110000 & 0.34 &  cold core & Bianchi et al. (2003) & IRAM data \\
1200 & 66000 & 0.57 & H$_{2}$ (NGC 4565) & this work & \\
\hline
\end{tabular}
\label{tab_emissivity}
\end{table*}

\begin{figure*}[t]
\epsfig{figure=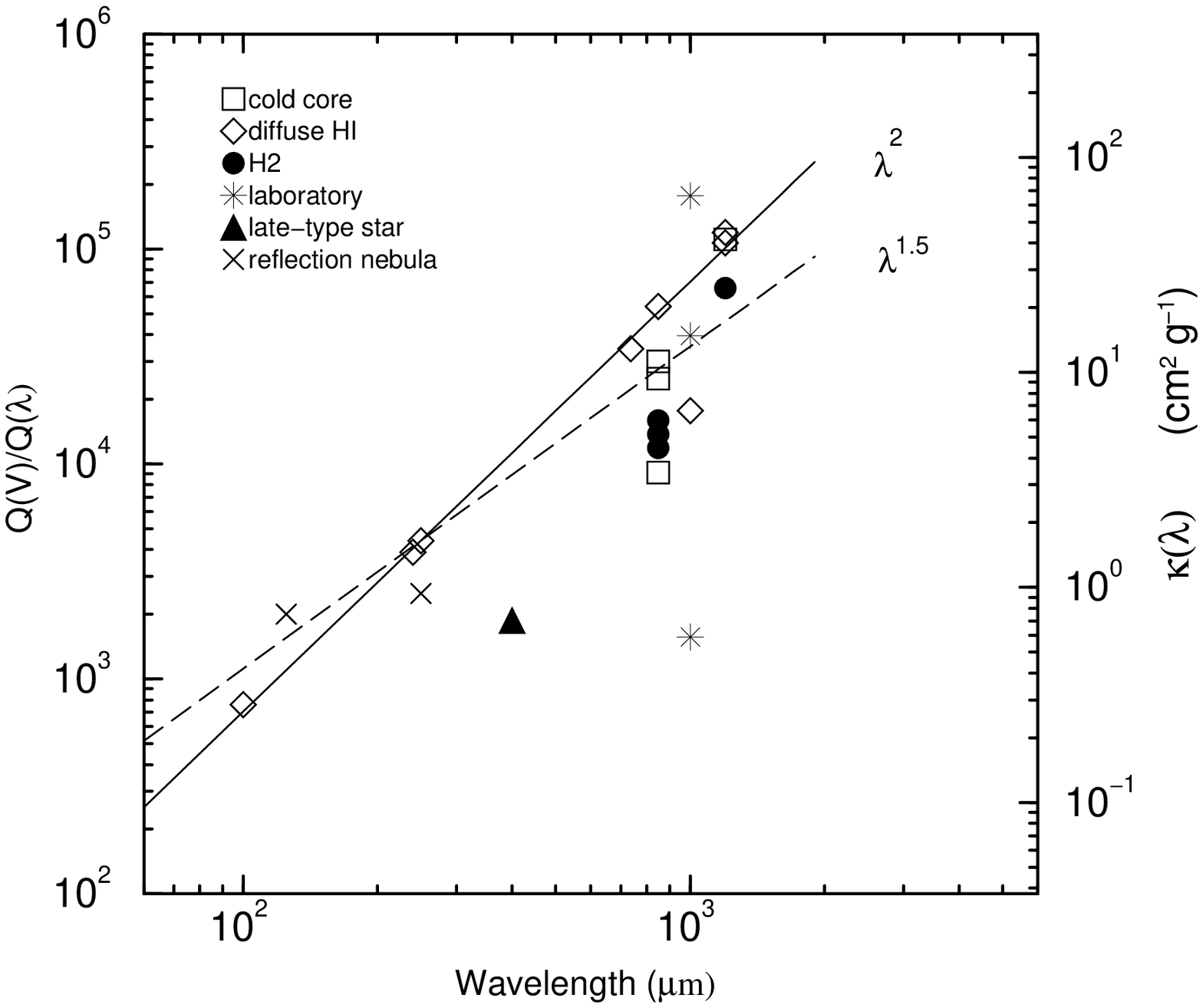,width=16.5cm}
\caption{Emissivity $Q(\lambda)$ of astrophysical dust in the far-infrared, submm and millimeter wavebands.
The plot depicts the values collated in Table~\ref{tab_emissivity} employing identical markers for similar
astrophysical environments. The estimates that we derive for NGC 4013, NGC 4565, NGC 5907 and NGC 891 (the
latter in Alton et al. (2000)) have been attributed to dust residing in molecular gas clouds (solid circles).
The dashed and dotted lines correspond to a wavelength dependency for the emissivity which varies as
$\beta$=2.0 and $\beta$=1.5 respectively ($Q \propto \lambda^{-{\beta}}$). The lines have been set to pass
arbitrarily through the frequently-cited model value of Draine \& Lee (\cite{draine1984}) at $250~\mu$m.}
\label{fig_emissivity}
\end{figure*}

\section{Deriving the emissivity $Q$}
\label{emissivity}

In order to compare our two chosen tracers of galactic dust - visual optical depth and submm/mm flux density - on equal terms we are compelled to smooth the map of $\tau_{V}$ from the RT model to the same spatial resolution as the SCUBA/IRAM data from which we wish to determine the emissivity. For NGC 4013 and NGC 5907 our analysis was for the $850~\mu$m waveband whilst for NGC 4565 the comparison was conducted at 1.2 mm. We emphasize that the $850~\mu$m and 1.2 mm filters lie on the Rayleigh-Jeans tail of the SED which renders our comparison less susceptible to uncertainties in the adopted grain temperature. After smoothing the map in V-band optical depth to the required spatial resolution we profiled along the major axis of the galaxy in order to derive the mean ratio $\frac{\tau_{V}}{f(\lambda)}$ [required for Eq.~(\ref{eq_sb})]. These profiles are illustrated in Fig.~\ref{fig_prof}. Profiles were also generated {\em perpendicular} to the major axis and these indicate that visual extinction and $850~\mu$m surface brightness decrease in a similar manner for increasing z-height (distance above the midplane). We do not attach too much significance to the correlation in z-height because our spatial resolution ($16''$) is too coarse with respect to the expected exponential scaleheight of galactic dust (0.25 kpc or $\sim 5''$ for a galaxy 10 Mpc away).

The profiles in Fig.~\ref{fig_prof} indicate that the visual optical depth falls off radially in a manner similar to that of the submm/mm surface brightness. We have, therefore, some confidence that the two quantities are tracing the same component of galactic dust. Spikes in the submm/mm emission are evident which cannot be accounted for by the smooth dust distribution adopted by the RT model. We emphasize that such localized peaks, at least for NGC 4013, are real within the noise and, as such, might either correspond to concentrations of molecular gas or warmer thermal emission from grains situated within star-forming regions. A discussion of how the submm/mm emission relates to the various gas phases of the ISM is deferred to Sect.~\ref{gas}. The percentage of $850~\mu$m emission contained in the spikes of Fig.~\ref{fig_prof} can be gauged by fitting a smoothly-declining baseline to the profile and totalling the emission above this level. The method is somewhat crude but yields a percentage of $\sim$20\% for the locally-enhanced $850~\mu$m flux density in NGC 4013. We conclude that both the visual extinction and submm thermal emission appear to trace the same component of galactic dust but a significant fraction ($\sim \frac{1}{5}$) of the submm emission cannot be accounted for by our simplified model of visual extinction.

From the profiles in Fig.~\ref{fig_prof} we obtain mean values of $51\pm 2$ and $50\pm 2$
for the ratio $\frac{\tau_{V}}{f({850\mu}m)}$ in NGC 4013 and NGC 5907, respectively, and a value of $330\pm 40$ for 
the ratio $\frac{\tau_{V}}{f(1.2 mm)}$ in NGC 4565.  Here, $f(850~{\mu}m)$ and $f(1.2~mm)$ refer to the surface
brightness at $850{\mu}m$ and 1.2 mm in Jy/$16''$~beam and Jy/$20''$~beam respectively.
In Sect.~\ref{sed} we have decomposed the SED of each object into a warm and a cold dust component in order to determine
the grain temperature for the blackbody function. We now intend to use Eq.~(\ref{eq_sb}) to determine $Q(850~{\mu}m)$ and 
$Q(1.2~mm)$ but, before doing so, we subtract off the minor contribution ($<20$\%) of the warm dust component to 
$f({850\mu}m)$ and $f(1.2~mm)$. This allows us to substitute a single grain temperature into Eq.~(\ref{eq_sb}) 
namely that of the cold dust component listed in Table~\ref{tab_sed}. This simplified approach yielded a 
ratio $\frac{Q(V)}{Q(850{\mu}m)}$ of 14000 and 12000 for NGC 4013 and NGC 5907, respectively, and a ratio 
$\frac{Q(V)}{f(1.2mm)}$ = 66000 for NGC 4565. Assuming $Q(V)\simeq 1.5$ (Sect.~\ref{tech}), we infer an 
emissivity of $\simeq 1.2 \times 10^{-4}$ and $\simeq 2.3 \times 10^{-5}$ at $850~\mu$m and 1.2 mm, 
respectively. For our adopted grain parameters ($a=0.1~\mu$m and $\rho=3000~$kgm$^{-3}$) this equates to 3.0 and 
0.57 cm$^2$g$^{-1}$ for $\kappa(850~{\mu}m)$ and $\kappa(1.2~mm)$ respectively. 

\section{Discussion on the emissivity (Q)}
\label{discussion}

In Table \ref{tab_emissivity} we compare our results with existing estimates of submm/mm emissivity for a 
diversity of astrophysical media (diffuse HI clouds, reflection nebulae etc). The corresponding dust absorption 
coefficient $\kappa$ is given assuming a classical grain size of $0.1~\mu$m, $Q(V)=1.5$ and a material density
 typical of amorphous silicates $\rho = 3 \times 10^{3}~ kgm^{-3}$ (Mennella et al. \cite{mennella1998}). The 
 same data are presented graphically in Fig.~\ref{fig_emissivity} with identical markers denoting similar grain
  environments. Most of the points in Fig.~\ref{fig_emissivity} rely, as we do in this paper, on observed 
  optical/NIR extinction to quantify the amount of dust emitting in the submm/mm regime. Notable exceptions are 
  the semi-empirical model of Draine \& Lee (1984) and the recent analysis of James et al. (2002). The former, 
  already alluded to in Sect.~\ref{intro}, constitutes a benchmark study which is frequently cited in the 
  literature (see also Li \& Draine (2001), a recent update of the model which proposes only modestly differing 
  emissivities in the FIR). James et al. (2002) derives the emissivity by quantifying the amount of dust on the
   basis of local gas density and local metal abundance (a constant fraction of metals is assumed to be tied up 
   in dust grains). In a similar manner, estimates by Dumke et al. (1997) and Neininger et al. (1996) for
    NGC 5907 and NGC 4565, respectively, are based on the column density of atomic hydrogen detected in the 
    outer regions of the galaxy. A solar gas-to-dust ratio is employed in order to quantify the amount of dust 
    present and the level of 1.2 mm emission is compared with this quantity of dust in order to infer the 
    emissivity.

The values derived in this paper, plus that of NGC 891 based on the same technique, indicate a low ratio 
$\frac{Q(V)}{Q(FIR)}$ and therefore a {\em high} efficiency for submm/mm thermal emission compared with, 
for example, COBE measurements of high-lattitude dust in the Milky Way (difference of factor 4 at $850\mu$m). 
Indeed, our estimates are more aligned with denser gas environments, where grains are expected to coagulate into 
amorphous particulates rather than remain as well-ordered crystalline cores. Our results lie within the range 
suggested by recent tests on {\em amorphous} carbon and silicate particulates conducted in the laboratory. 
Submm/mm emissivity is expected to be environment sensitive responding directly to grain temperature and, 
indirectly (through coagulation), to ambient gas density. Recent observational evidence points to a threefold 
increase in $Q(600~{\mu}m)$, compared to the diffuse ISM, for fluffy, coagulated grains forming in dense dust
 clouds (Stepnik et al. \cite{stepnik2003}). Experiments in the laboratory on amorphous grains predict a fall 
 of factor 2-6 in $Q(\lambda \sim 1~mm)$ as the grain temperature lowers from 300 K to 24 K. Results presented 
 below indicate a strong association between $850\mu$m emission and the $^{12}$CO(1-0) line in our sample of 
 galaxies. Gas clouds in the Milky Way that emit strongly in the CO-line are characterised by densities of
  $\sim 10^{3}~$m$^{-3}$ (Seaquist et al. 2004) and are thus more akin to dense clouds rather than the diffuse
   HI clouds analysed in the aforementioned COBE study. James et al. (2002) claim to derive the same emissivity
    regardless of the value of $X$ used to convert the $^{12}$CO(1-0) line emission of their sample into 
    H$_{2}$ gas masses.  This seems difficult to explain unless their sample is dominated by HI-rich, metal 
    poor objects rather than the H$_{2}$-rich galaxies that make up our own sample. 

Before accepting the possibility that the relatively high emissivity values we derive are attributable to grains 
in denser H$_{2}$ clouds, we carry out a critical examination of the assumptions and modalities of our 
technique. These can be summarized as follows: (i) the grains responsible for optical/NIR extinction are 
{\it not}, as supposed, the origin of the submm/mm emission, (ii) our SED fitting is erroneous implying an 
incorrect grain temperature (iii) we {\it overestimate} the submm/mm flux densities (iv) we {\it underestimate} 
the visual opacity. We discuss each of these points in turn below.

\subsection{Error in the basic assumptions}
\label{assumption}

Our central assumption is that grains extinguishing optical/NIR radiation in spiral disks will also be the main 
emitters of submm/mm thermal radiation. Hildebrand (1983) has already shown that the dependencies of optical 
extinction and FIR emission on grain size are broadly similar. We make a critical appraisal of this argument in
the Appendix using the MRN grain-size distribution as a test-case (Mathis et al. 1977). To summarize the
conclusions of that section, we find that, for the same grain-size distribution in all parts of the ISM,
grains of roughly the same size will dominate both optical extinction and submm/mm thermal emission. The enhanced
emissivity we measure in the previous section, however, may suggest that our SCUBA images are sensitive to a population 
of grains that are significantly larger than the classical $0.1 \mu$m grains dominating optical extinction. This
question of bi-modality in the grain size can only be solved expeditiously by a decomposition of
galactic submm/mm emission into discrete and diffuse sources (as has been carried out for the Milky Way, at shorter 
wavelengths, using IRAS). A project of this kind is probably beyond the reach of current submm/mm arrays
where the field of view ($2-3'$) and, therefore, ill-suited to a large-scale survey of the Galactic plane.
Prior subtraction of discrete sources from our SCUBA images, in order to eliminate larger, non-classical
grains, would act to {\em lower} the emissivity we derive for the classical population dominating the optical
extinction.
 
\subsection{Uncertainties in SED fitting}
\label{sed_uncertainties}
SED fitting is rendering particularly uncertain by the value chosen for wavelength-dependency $\beta$. SED fits with $\beta=2$, rather than the $\beta=1.5$ adopted in Sect.~\ref{sed}, imply {\em lower} grain temperatures and therefore {\em higher} submm/mm emissivities. For example, we derive 13 K for the cold dust component in NGC 4013 (instead of 23 K with $\beta=1.5$) and consequently $\frac{Q(V)}{Q(850{\mu}m)} = 6500$ (i.e. a submm emissivity which is twice as high). In conclusion, adopting values of $\beta$ higher than 1.5 would increase the divergence of our results from COBE-based estimates of emissivity and the Draine \& Lee (\cite{draine1984}) model.

Non-equilibrium emission from very small grains and PAHs is also expected to contribute to the flux density detected in the IRAS filters. Consequently we subtract 62\% and 14\% from the flux densities at 60 and $100~\mu$m (Desert et al. 1990) and re-fit the modified SED to gauge the magnitude of this effect. The implied change in the emissivity is $\sim 10$\%. In summary, since the $850~\mu$m filter samples emission on the Rayleigh-Jeans tail, the emissivity we infer at that wavelength is only linearly-proportional to the uncertainties in the grain temperatures. Our lattitude in fitting $T$ is too small to account for the relatively high emissivities we derive.

\subsection{An overestimate of submm/mm emission ?}
\label{overestimate}

The SCUBA filter at $850~\mu$m and the IRAM continuum bandpass at 1.2 mm are expected to contain some line emission 
from the respective transitions $^{12}$CO(3-2) and $^{12}$CO(1-0). For the former, contamination is expected to be as
 high as 50\% for dense gas situated within 1-2 kiloparsec of the galactic nucleus (Bianchi et al. \cite{bianchi2000}).
  Similarly, Seaquist et al. (2003) estimate a mean contamination of 25\% for the infrared-luminous galaxies making up 
  the SCUBA Local Universe Survey (SLUGS; Dunne et al. \cite{dunne2000}). For the general disk, however 
  (appropriate to the analysis we are carrying out here), the CO gas is expected to be less collisionally excited and 
  the corresponding $^{12}$CO(3-2) contamination is believed to be much closer to 10\% (Israel et al. \cite{israel1999}; 
  Boettner et al. \cite{boettner2003}). Images of NGC 5907 in the $^{12}$CO(3-2) emission line (Dumke et al in prep.)
  indicate a contamination of 130mJy to the SCUBA continuum filter (i.e.~less than 10\% of the continuum level). 
  
  IRAM heterodyne and continuum measurements of nearby, quiescent disks imply corrections of $\sim$ 10\% are necessary to
  eliminate the $^{12}$CO(1-0) line emission from the 1.2 mm continuum filter (e.g.~Guelin et al. 1993). The contribution 
  of the $^{12}$CO(6-5) line to the SCUBA filter 
  at $450~\mu$m is expected to be less than 10\% (Seaquist et al. 1994). To gauge the impact of non-thermal emission 
  on our results we have re-fitted the SEDs in Fig.~\ref{fig_sed} removing, beforehand, both non-equilibrium emission 
  from small grains (Sect.~\ref{sed_uncertainties}) and a line-contamination of 10\%, 20\% and 10\% to the wavebands 
  $450~\mu$m, $850~\mu$m and 1.2 mm respectively. The impact on our infered values of $\frac{Q(V)}{Q(850{\mu}m)}$ is 
  comparatively small ($\sim$15\%) and cannot account for the relatively high submm/mm emissivities which we have found.

\begin{figure}
\epsfig{figure=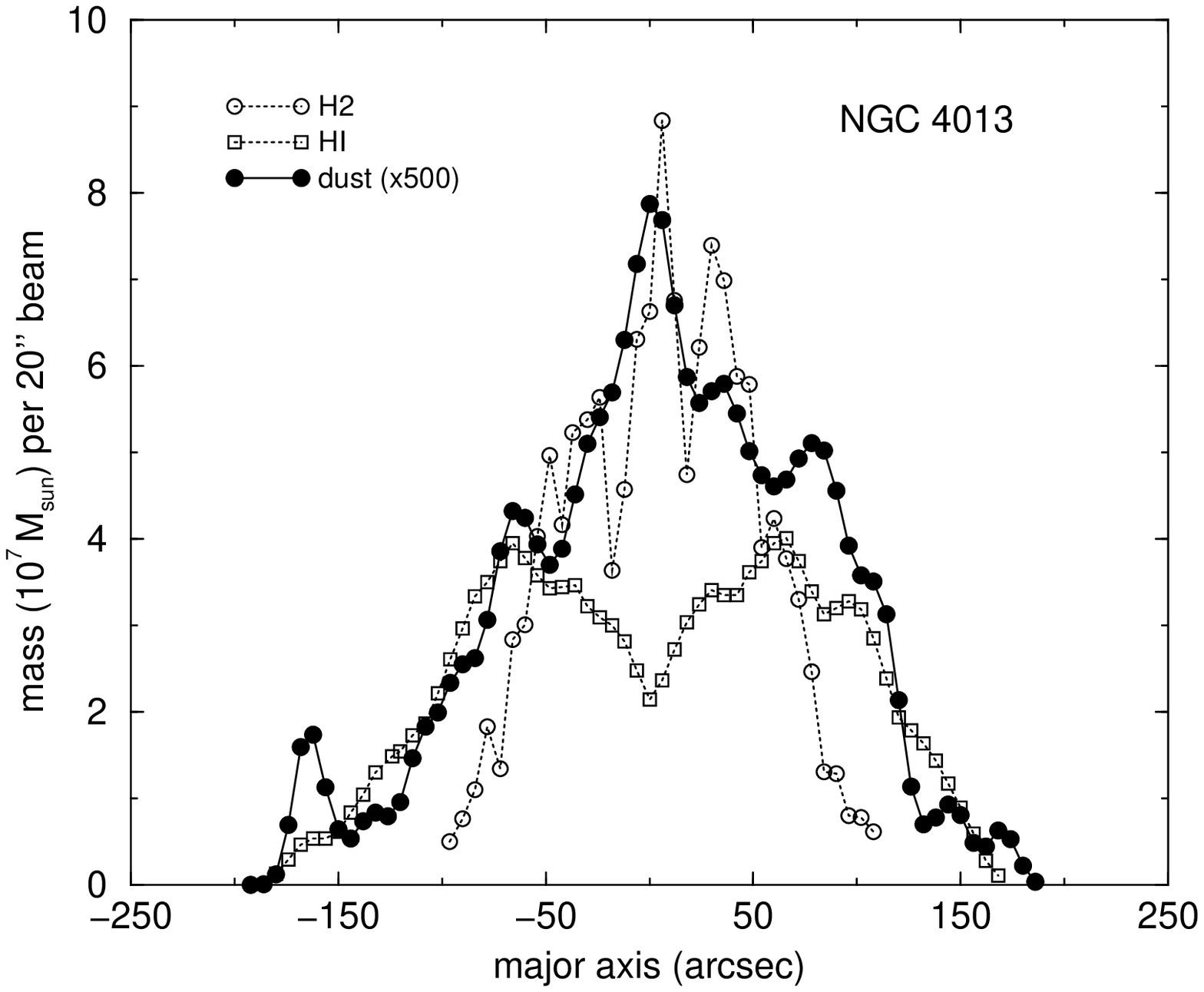,width=8.5cm}
\epsfig{figure=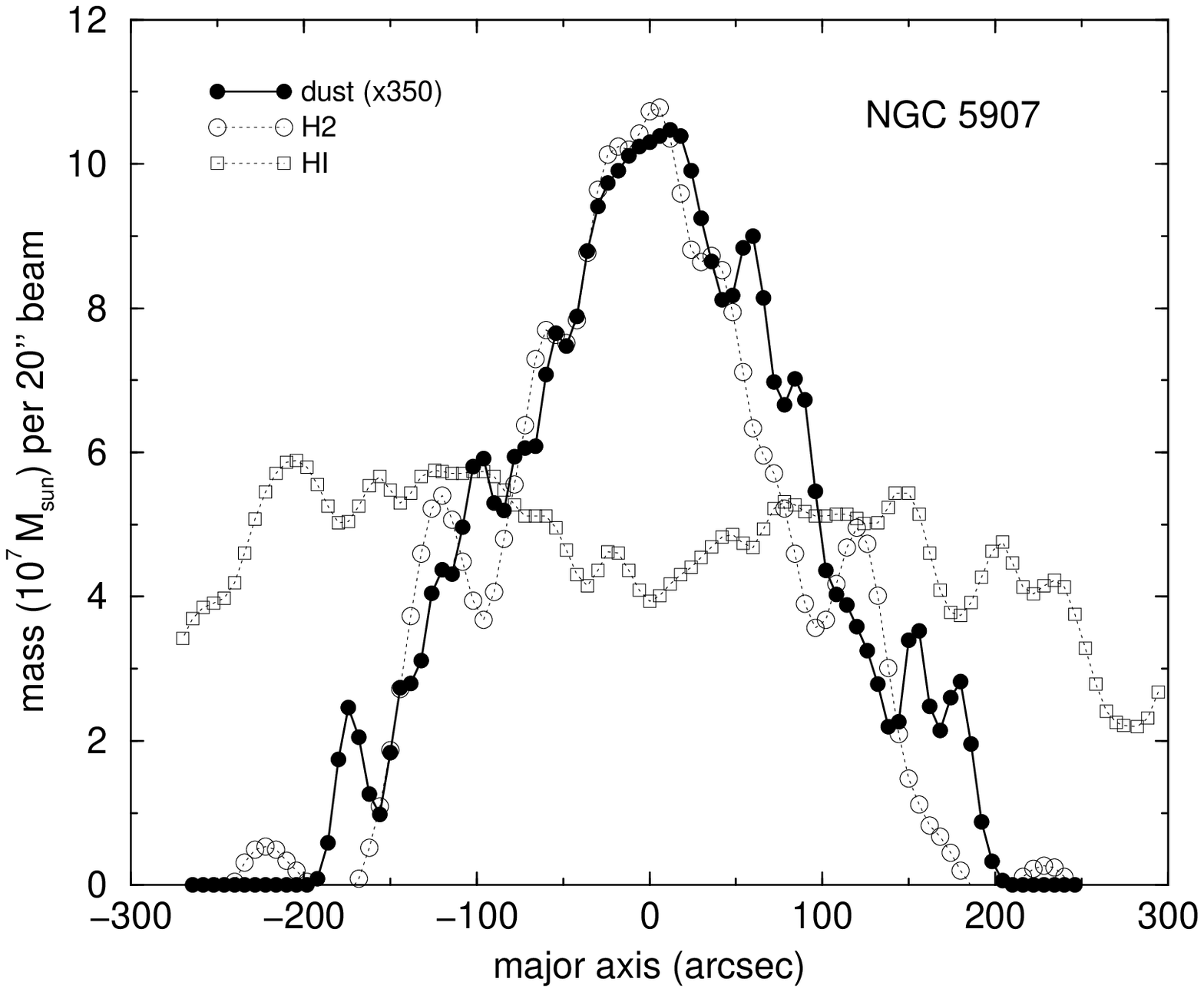,width=8.5cm}
\caption{Dust and neutral gas profiles along the major axis of NGC 4013 and NGC
5907. The solid circles
denote the distribution of grains (using multiplicative factors of 500 and 350 to clarify the plot).
 The open circles and open squares trace the respective distributions of molecular and atomic gas
within the disk.
The major axis is defined such that positive values correspond to north-east and south-east
for NGC 4013 and NGC 5907 respectively}
\label{fig_gas_profile}
\end{figure}

\subsection{An underestimate of optical opacity ?}
\label{underestimate}

In a synopsis of disk opacity, Alton et al. (\cite{alton2001}) found that the optical depth implied by the RT-model of Xilouris et al. (\cite{xilouris1999}) was generally low compared to other values cited in the literature. Many of these other studies, however, are based on far-infrared emission (e.g.~ISOPHOT maps at $200\mu$m) and assume {\em a priori} an emissivity which is lower than that derived here (implying necessarily a relatively high visual optical depth). This was the conclusion of Popescu et al. (\cite{popescu2000}) and Misiriotis et al. (\cite{misiriotis2001}), who, using the same RT model as us but employing the lower emissivity of Draine \& Lee (\cite{draine1984}), explain the detected submm/mm emission from external galaxies in terms of a second, massive dust component distributed `thinly' in z-height in a similar fashion to the OB stellar population. The `thin' nature of this disk is imperative - increasing the optical depth of the `standard' dust disk we have assumed here (scale-height of dust half that of the stars) would lead to a flagrant discrepancy between the model results and the observed photometry. We ask ourselves whether this second dust disk, which would increase the visual optical depth by a factor of 3, is realistic. IRAS and COBE (DIRBE) images of the solar neighborhood indicate a dust scaleheight of 0.13 kpc with the possibility of cold dust having an exponential scaleheight as great 0.5 kpc (Davies et al.~\cite{davies1997}). This is larger than the 0.09 kpc scaleheight characterising the OB stellar population within the Galaxy. We admit that we cannot rule out the presence of a second, massive dust with our current observations and that this must be seriously considered as an alternative to the relatively high emissivity we have derived. Essentially, with a massive thin disk, the near-infrared becomes optically thick and the edge-on surface photometry is no longer diagnostic. Moreover, the thickness of the second dust disk approaches that of the NIR spatial resolution.  
A discriminatory test will be to predict the azimuthally-averaged colours of face-on spiral disks (e.g. B-K) using (i) an RT-model containing a massive, thin dust layer (low emissivity) and (ii) an RT-model with only a `standard' dust layer (high emissivity). A comparison between model and observation is being undertaken by one of us as a thesis project (Dasyra) but no conclusions from this study are currently available.

A component of dust in clumps too optically thick and compact to be probed by RT simulations is feasible. However, as argued by both Alton et al. (\cite{alton2000}) and Misiriotis \& Bianchi (\cite{misiriotis2002}), this is unlikely to exceed 50\% of the mass assumed in the current, smooth dust distribution. A correction of the magnitude is insufficient to explain our results.

\section{The submm as a tracer of gas}
\label{gas}

The dust in our RT-models has been assumed to be smoothly distributed in order to simplify our calculations. 
The major axis profiles plotted in Fig.~\ref{fig_prof} indicate, however, that a conspicuous fraction of the
 $850~\mu$m thermal emission detected within our sample is clumped at the kpc-level. NGC 4013 emits, 
for example, much more copiously in the north-east part of the disk (analogous to the enhanced submm emission
 detected in the north-east disk of NGC 891; Alton et al. \cite{alton2000}). We have argued that a possible 
physical interpretation of the high emissivities derived in Sect.~\ref{discussion} lies in an association between 
submm/mm emission and concentrations of {\em molecular gas}. In this section we examine this hypothesis further 
and compare the submm/mm emission within our sample with the corresponding distributions of molecular 
and atomic gas within the disk. In this way we hope to learn which medium within the disk is harbouring the 
grains giving rise to the submm/mm emission. In the past, tight spatial correlations have been established 
between CO line emission and the $450~\mu$m, $850~\mu$m and 1.2 mm thermal emission detected from nearby, 
quiescent disks such as NGC 7331, NGC 891 and NGC 6946 (Guelin et al \cite{guelin1993}; 
Alton et al. \cite{alton2000}; Bianchi et al. \cite{bianchi2000}; Alton et al. \cite{alton2001}).

Information regarding the distribution of gas along the major axis was obtained from the following sources. For
NGC 4013, both the atomic and molecular hydrogen column densities were extracted from Fig. 5 of 
Gomez de Castro \& Garcia-Burillo (1997). For NGC 5907, the $^{12}$CO(1-0) line intensity and 21 cm emission were
taken from Fig. 6 of Dumke et al. (1997) using their Fig. 5 to calibrate the latter. For an indication of how dust and gas are correlated within NGC 4565 we refered to the existing study of Neininger et al. (1996). To convert the $^{12}$CO(1-0) emission to a H$_{2}$ column density we adopted a value of X= 2 $\times 10^{20}$ cm$^{-2}$ K km/s (likewise, values of H$_{2}$
column density for NGC 4013 from Gomez de Castro \& Garcia-Burillo (1997) were adjusted to this value of $X$). We
recognise that there is some debate concerning the exact value of $X$ (estimates vary from about 1.5 to 6
$\times 10^{20}$ cm$^{-2}$ K km/s for large, quiescent spiral disks) but values close to 
2 $\times 10^{20}$ cm$^{-2}$ K km/s seem appropriate for the general ISM of the Milky Way (Maloney 1990). The 
SCUBA $850~\mu$m images of NGC 4013 and NGC 5907 were smoothed to the same spatial resolution as the gas data ($20''$) 
before comparing the distribution of submm/mm emission with the gas.

Fig.~\ref{fig_gas_profile} illustrates the major axis distributions of gas and dust in NGC 4013 and NGC 5907.
The surface mass density for the dust follows from the parameter $n$ in Eq.~(\ref{eq_fd}) assuming the emissivity 
$Q$ we have derived in Sect.~\ref{emissivity} and a respective grain radius and material density of $0.1~\mu$m 
and 3000 kg$m^{-3}$. The profiles in Fig.~\ref{fig_gas_profile} reinforce the idea that grains emitting in the 
submm are associated with molecular gas, particularly within the central 10 kpc of the disk. Moving towards 
the disk edge,the fall-off in submm emission is intermediate between that of the molecular and atomic gas 
for NGC 4013. Neininger et al. (1996) recognise a similar trend for NGC 4565 at 1.2 mm and analyses of the submm and 
millimeter emission from NGC 891 present similar radial correlations (Alton et al. 2000; Guelin et al. \cite{guelin1993}).
For NGC 5907, the $850~\mu$m appears to follow the molecular gas at all radii and falls to zero when only 
atomic gas is present.

The tight correlations we observe between $850~\mu$m emission and the $^{12}$CO(1-0) line intensity might lead us to the 
conclusion that the SCUBA longwave filter is contaminated much more severely by $^{12}$CO(3-2) line emission than the 
$\leq 20$\% we have argued in Sect.~\ref{overestimate}. We emphasize, however, that correlations between
CO-line emission and submm/mm thermal emission have already been noted when the continuum filter in use is known to
contain only minor line contamination. For example, a strong association
between 1.2 mm continuum emission and $^{12}$CO(1-0) line intensity has already been established in both NGC 891 
(Guelin et al. \cite{guelin1993}) and NGC 5907 (Dumke et al. \cite{dumke1997}). These observations, along with the
estimates of CO-line contamination produced in Sect.~\ref{overestimate}, persuade us that the correlation we find
between $850~\mu$m surface brightness and $^{12}$CO(1-0) line emission indicate a true physical association 
between CO gas and cold dust.

\begin{figure}
\epsfig{figure=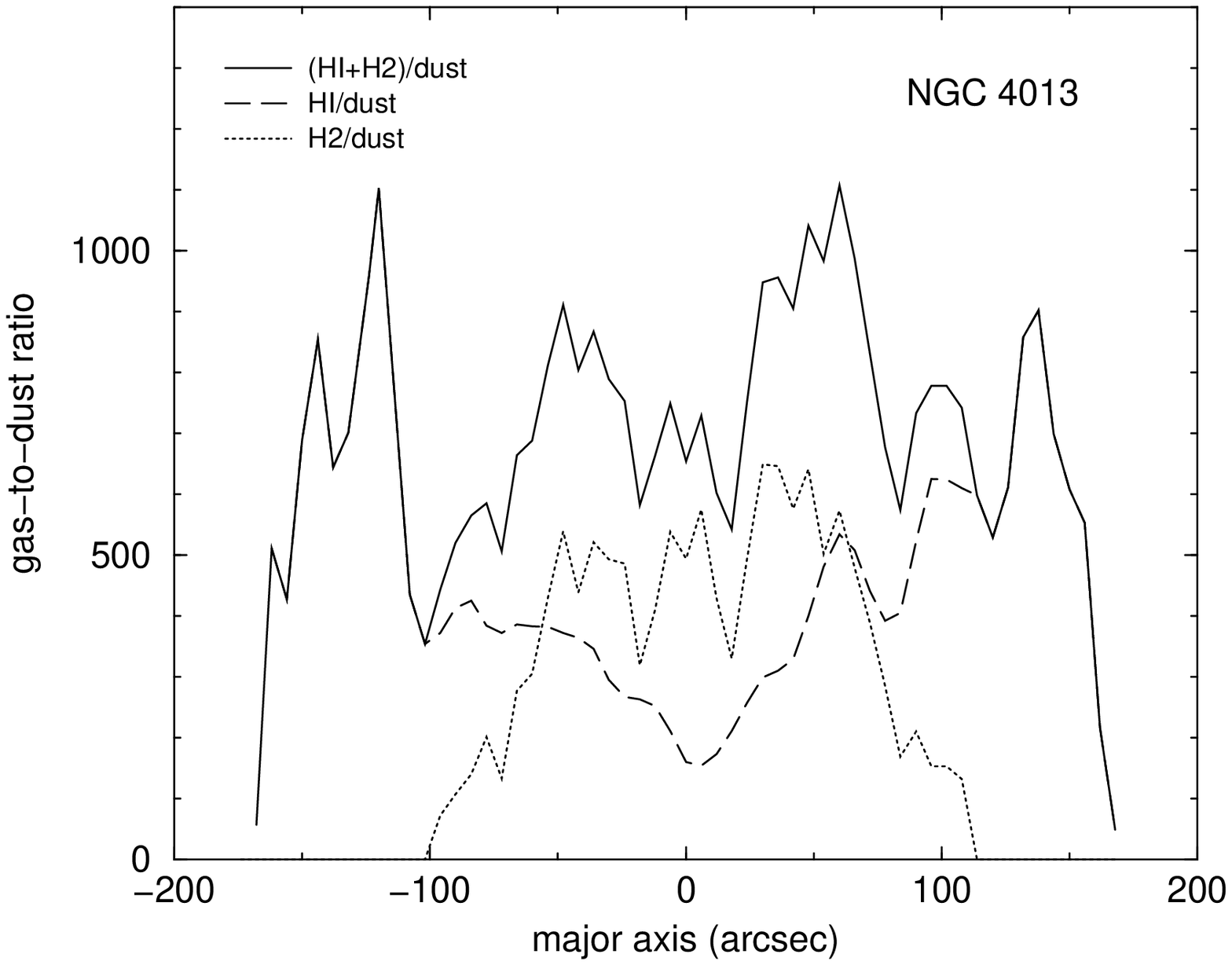,width=8.5cm}
\epsfig{figure=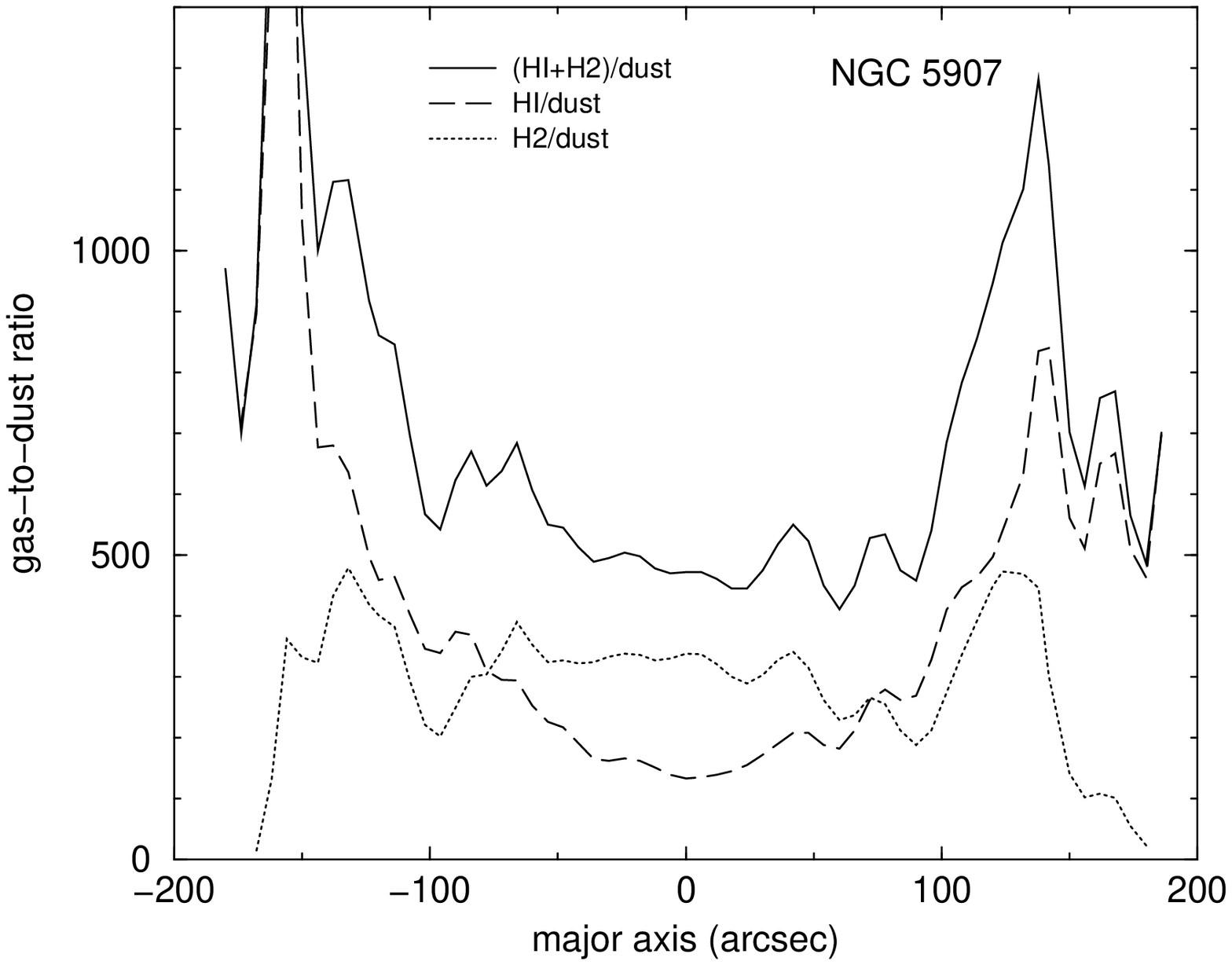,width=8.5cm}
\caption{The gas-to-dust ratio along the major axes of NGC 4013 and NGC 5907.
 The CO line has been converted to a H$_{2}$ column density assuming $X= 2 \times 10^{20}$ cm$^{-2}$
K km/s. The dust column density follows from the submm emissivities derived in Sect.~\ref{emissivity} of
this work.}
\label{fig_dustgas_profile}
\end{figure}

We use our profiles of gas and dust surface density along the major axis to infer the gas-to-dust
ratio. This is illustrated in Fig.~\ref{fig_dustgas_profile} for NGC 4013 and NGC 5907. Furthermore, we
employ the trend noted by Garnet (\cite{garnet1988}) between absolute B-band magnitude and relative oxygen abundance
in order to infer the gas-phase metallicity (see also Garnett \& Shields \cite{garnet1988}). 
Although extinction effects are severe in these edge-on disks, our
radiation transfer model can be used to derive the intrinsic blue luminosity of NGC 4013 and NGC 5907. The object
in question can then be located on the O/H vs.~$M_{B}$ plot of Garnet. Following Garnet, we adopt 
$H_{0}$=50 kms$^{-1}$/Mpc in order to determine the distance modulus and hence absolute B-band magnitude.
 In this way, we derive values of -19.8 and -20.5 for $M_{B}$ in NGC 4013 and NGC 5907, respectively.
 On the Garnet plot this corresponds to a metallicity of $z=0.79z_{\odot}$ (for NGC 4013) and $z=z_{\odot}$ (for 
NGC 5907) i.e. a solar-type metallicity within 
the 40\% uncertainties inherent in the estimate.

The error-weighted gas-to-dust ratios that we infer from Fig.~\ref{fig_dustgas_profile} are close to 500 and,
therefore, a factor 2-3 higher than the values of 150-300 associated with the solar neighborhood.
We emphasize that our gas-to-dust estimates are
sensitive to the chosen value of the conversion factor for $^{12}$CO(1-0) to H$_{2}$. 
Frequently, $X\simeq 0.5 \times 10^{20}$ cm$^{-2}$ K km/s has been proposed for the collisionally-excited 
climes characterising the central few kpc of spiral galaxies (e.g. Neininger et al. \cite{neininger1996}). Reducing 
our $X$ parameter to this level would bring our gas-to-dust ratio much closer to the solar value.

\section{Summary and conclusions}
\label{summary}

We have reduced and collated submillimeter and millimeter imaging data for three nearby edge-on spiral galaxies (NGC 4013,
NGC 4565 and NGC 5907). We have compared the distribution and level of thermal continuum emission from
the dust with the V-band optical depth inferred from a sophisticated (scattering+absorption) radiation
transfer simulation (Xilouris et al. \cite{xilouris1999}). The analysis yields the emissivity of dust at wavelengths 
of $850~\mu$m and 1.2 mm. At $850~\mu$m we infer an emissivity of $1.2 \times 10^{4}$ i.e.~a factor 4 {\em higher} than 
both predictions from the Draine \& Lee (\cite{draine1984}) model and measurements made on high-lattitude Milky Way dust 
using COBE. At 1.2 mm, our estimate is only a factor 1.5 higher. Our results, particulary at $850\mu$m, are 
consistent with recent measurements carried out under laboratory conditions where
relatively high emissivity values are recorded for amorphous silicate and carbon particulates. The submm/mm
emission detected from our objects follows closely the distribution of molecular gas, the latter being traced 
in the $^{12}$CO(1-0) line. A physical explanation for the relatively high emissivity values we derive might
 be that dust emitting strongly in the submm/mm waveband is situated chiefly in molecular gas clouds where the
elevated density is conducive to the formation of amorphous, fluffy grains. Such grains are expected to possess emissivity 
values that are a factor 3 or so higher than refactory cores circulating in the diffuse ISM (Ossenkopf \& Henning 
\cite{ossenkopf}).

We accept that our radiation transfer simulation could `miss' a significant fraction of galactic dust
{\em if the grains are distributed within a non-standard way within the disk} (i.e. within a very narrow layer
with respect to the mid-plane). A `litmus test' for both the radiation transfer model, and the high emissivities
we derive as a consequence, will be a prediction of the observed optical and near-infrared azimuthally-averaged
radial colours of face-on spirals assuming the optical depths we use in this work. For example, an observed B-K colour 
which is higher than that predicted will indicate an underestimate of visual optical depth for the current edge-on sample.
Work is currently being undertaken by one of us (Dasyra, thesis project) to resolve this issue. 

We also accept that a 
major part of the submm/mm thermal emission detected in our SCUBA/IRAM images may arise from
bigger grains contained within localised, compact sources (e.g. pre-stellar cores or circumstellar disks) whilst the
optical extinction we measure is almost certaintly attibutable to `classical' grains (characterised by a Galactic
extinction law). If a bimodality of this kind is indeed present in our observations, we would also overestimate 
the emissivity associated with diffuse interstellar dust. 

The gas-to-dust ratios we derive for NGC 4013 and NGC 5907 are $\simeq 500$ and therefore a factor 2-3 
higher than values cited for the solar neighborhood. Our estimates depend sensitively on the adopted value
of $X$, the conversion of $^{12}$CO(1-0) line intensity to H$_{2}$ column density. We have used 
$X=2 \times 10^{20}$cm$^{-2}$ K km/s in this work but, if we were to follow proponents of 
$X=0.5 \times 10^{20}$cm$^{-2}$ K km/s, our gas-to-dust estimates would lie close to solar values.
A crude estimate of metallicity using the $M_{B}$-O/H relation (Garnet \cite{garnet1988}) suggests a
metallicity close to solar for NGC 4013 and NGC 4565.

\section{Appendix}
\label{appendix}
Here, we make a critical appraisal of the central supposition in this paper, namely that the same grains
responsible for optical extinction also give rise to the submm/mm thermal continuum.
The mean radius ($a$) of grains contributing to optical extinction $\tau_{opt}$ can be expressed as follows: 

\begin{equation}
\label{equ_mean}
<a> = \frac{\int a n(a) \tau_{opt} da}{\int n(a) \tau_{opt} da}
\end{equation}

\noindent where the integral is weighted by the distribution of grains sizes $n(a)$ (e.g. Hildebrand 
\cite{hildebrand1983}). The linear behaviour of the Galactic extinction law within the optical regime 
indicates that $Q_{opt} \propto \frac{a}{\lambda}$ (Whittet \cite{whittet1992}). Refering to the definition 
of optical depth [Eq.~(\ref{eq_tau})], Eq.~(\ref{equ_mean}) then becomes:

\begin{equation}
\label{equ_mean_q}
<a> = \frac{\int a n(a) a^{3} da}{\int n(a) a^{3} da}
\end{equation}

We use the MRN size distribution as a test case where $n(a) \propto a^{-3.5}$ over the range 
$0.01~{\mu}m \ge a \ge0.25~{\mu}m$ (Mathis et al. \cite{mathis1977}). This yields a weighted grain radius of 
$0.1~\mu$m corresponding to the frequently-cited `classical' grain size. We emphasize however 
that, {\em within the limits set out below}, the exact 
size distribution is usually not important because emission in the submm/mm can also be shown to
behave as Eq.~(\ref{equ_mean_q}). For wavelengths much larger than the grain size, extinction is dominated by alsorption 
and the Kramers-Kronig relations yields the approximation $Q \propto a$ in the submm/mm regime (Whittet 1992).
 It is the same parameter $Q$ which determines the efficiency with which grains emit submm radiation. Thus, for a
 single grain population characterised by an MRN size distribution, we expect grains of size $0.1\mu$m to
 dominate both optical extinction and submm/mm thermal emission.
 
The arguments advanced above are less robust if grains of different sizes are believed to dominate different parts
of the ISM. Thus, we can envisage a scenario whereby optical extinction arises from a widely-distributed
population of classical dust grains whereas a significant fraction of submm/mm thermal emission emanates from
modified, larger grains situated in dense, cold gas regions. There is some evidence that a bimodality of this kind
has been detected in the present study. The optical/NIR extinction law within our sample of galaxies is very similar
to that of the Milky Way (Sect.~\ref{tau_v}) indicating an absolute-to-selective extinction ratio of $R_{V}\simeq 3$.
In contrast, the emissivity we measure in the submm/mm is relatively high pointing to grains of either an
amorphous structure or larger size. There is some evidence that dark dust clouds, possessing
elevated submm/mm emissivities, are also characterised by higher total-to-selective extinction ratios
($R_{V}=4-6)$ and hence larger grain sizes (Kandori et al. 2003).

Bimodality would not normally be expected to be a severe problem in this study if the two grain populations in
question are well mixed and uniformly distributed with respect to the heating sources.
Eq.~(\ref{equ_mean_q}) tells us that the larger grains will tend to dominate both optical extinction and submm/mm
thermal emission. It is known, however, that observations of galactic-scale extinction are weighted towards
widely-distributed, diffuse dust clouds rather than localised, dust clumps (Misiriotis \& Bianchi 2002). In
contrast, larger grains appear to be restricted to compact media such as circumstellar shells or pre-stellar cold
cores. Given the temperature of 10-20 K we measure for the cold dust in this study (Sect.~\ref{sed}), we rule out the
 environs of young stars as the primary source of compact submm/mm thermal emission within our objects.
 
 To the best of our knowledge no current astronomical dataset exists from which we can surmise the proportion of 
 submm/mm emission emanating from discrete, localised sources within the Galactic plane. Such a study would allow
 us to gauge the proportion of submm/mm thermal emission arising from compact sources such as pre-stellar cores
 or circumstellar shells within our targets. Utilising COBE data for the Milky Way, Sodroski et al. 
(\cite{sodroski}) found that $\frac{2}{3}$ of 140 and $240~\mu$m emission comes from diffuse HI clouds and is, 
therefore, unlikely to be associated with discrete, compact sources. A comparable decomposition of sources at 
longer wavelengths has not been attempted due to the small field of view of current submm/mm arrays ($2-3'$ for SCUBA
and IRAM). This is ill-suited to a large-scale survey of the Galactic plane.

One further consideration is whether grains {\em of the same composition} are responsible for both optical 
extinction and submm/mm thermal emission. Most grain models that have been proposed are composed chiefly of 
a silicate and carbon/graphite component (Whittet \cite{whittet1992}). The carbon/graphite grains are 
expected to be more efficient at absorbing optical radiation (Draine \& Lee \cite{draine1984}). They will, 
however, also reach a higher equilibrium grain temperature and thus contribute more emission to the submm/mm 
waveband. Temperatures of 15-20 K and 10-15 K, respectively, are expected for carbon and silicate grains heated by
the general interstellar radiation field (Mathis et al 1983).


\begin{acknowledgements}
This research has made use of the {\it NASA/IPAC Extragalactic Database (NED)} which is operated by the
Jet Propulsion Laboratory, California Institute of Technology, under contract with the National
Aeronautics and Space Administration. We are indebted to the referee, Dr R. Laureijs, for pointing out the
association between enhanced absolute-to-selective extinction ratios and increased far-infrared emissivity.
\end{acknowledgements}


\end{document}